\documentclass[longauth]{aa}  %bibnumber
\usepackage{color}
\usepackage[colorlinks,breaklinks]{hyperref}
\hypersetup{linkcolor=blue,citecolor=blue,filecolor=black,urlcolor=blue}
\usepackage{amssymb}

\usepackage{xspace}

\usepackage{graphicx}
\usepackage{txfonts}
\usepackage{orcidlink}

\usepackage[switch]{lineno}
\begin{document}

%
%%%%%%%%%%%%%%%%%%%%%%%%%%%%%%%%%%%%%%%%
%%%%%%%%%%%%%%%%%%%%%%%%%%%%%%%%%%%%%%%%
%\usepackage[options]{hyperref}
% To add links in your PDF file, use the package "hyperref"
% with options according to your LaTeX or PDFLaTeX drivers.
%

   \title{ZTF SN~Ia DR2: Overview}
   \author{
    Rigault, M.\inst{\ref{ip2i}}\fnmsep\thanks{\texttt{m.rigault@ip2i.in2p3.fr}} \orcidlink{0000-0002-8121-2560},
    Smith, M.\inst{\ref{ip2i},\ref{lancaster}}
    \fnmsep\thanks{\texttt{mat.smith@lancaster.ac.uk}}
    \orcidlink{0000-0002-3321-1432},
    % -- %
    Goobar, A.\inst{\ref{okc}},\orcidlink{0000-0002-4163-4996},
    Maguire, K.\inst{\ref{dublin}}\orcidlink{0000-0002-9770-3508},
    Dimitriadis, G.\inst{\ref{dublin}}\orcidlink{0000-0001-9494-179X},
    Johansson, J.\inst{\ref{okc}}\orcidlink{0000-0001-5975-290X},
    Nordin, J.\inst{\ref{humbolt}}\orcidlink{0000-0001-8342-6274},
    % -- %
    Burgaz, U.\inst{\ref{dublin}}\orcidlink{0000-0003-0126-3999}, 
    Dhawan, S.\inst{\ref{cambridge}}\orcidlink{0000-0002-2376-6979}, 
    Sollerman, J.\inst{\ref{okc2}}\orcidlink{0000-0003-1546-6615},
    Regnault, N.\inst{\ref{lpnhe}},\orcidlink{0000-0001-7029-7901},    
    Kowalski, M.\inst{\ref{desi},\ref{humbolt}}\orcidlink{0000-0001-8594-8666},    
    Nugent, P.\inst{\ref{lbnl},\ref{ucb}}\orcidlink{0000-0002-3389-0586},    
    % -- %
    Andreoni, I.\inst{\ref{collegepark},\ref{collegepark2},\ref{goddard}}\orcidlink{0000-0002-8977-1498},
    Amenouche, M.\inst{\ref{canada}}\orcidlink{0009-0006-7454-3579},
    Aubert, M.\inst{\ref{clermont}},
    Barjou-Delayre, C.\inst{\ref{clermont}},
    Bautista, J.\inst{\ref{marseille}}\orcidlink{0000-0002-9885-3989},
    Bellm, E.\inst{\ref{dirac}}\orcidlink{0000-0001-8018-5348},
    Betoule, M.\inst{\ref{lpnhe}}\orcidlink{0000-0003-0804-836X},
    Bloom, J.~S.\inst{\ref{ucb}, \ref{lbnl}}\orcidlink{0000-0002-7777-216X},
    Carreres, B.\inst{\ref{marseille},\ref{duke}}\orcidlink{0000-0002-7234-844X},
    Chen. T.~X.\inst{\ref{ipac}}\orcidlink{0000-0001-9152-6224},
    Copin, Y.\inst{\ref{ip2i}}\orcidlink{0000-0002-5317-7518},
    Deckers, M.\inst{\ref{dublin}}\orcidlink{0000-0001-8857-9843},
    de Jaeger, T.\inst{\ref{lpnhe}}\orcidlink{0000-0001-6069-1139},
    Feinstein, F.\inst{\ref{marseille}}\orcidlink{0000-0001-5548-3466},
    Fouchez, D.\inst{\ref{marseille}}\orcidlink{0000-0002-7496-3796},
    Fremling, C.\inst{\ref{caltech}}\orcidlink{0000-0002-4223-103X},
    Galbany, L.\inst{\ref{barcelona1},\ref{barcelona2}}\orcidlink{0000-0002-1296-6887}
    Ginolin, M.\inst{\ref{ip2i}} \orcidlink{0009-0004-5311-9301},
    Graham, M.\inst{\ref{caltechphysics}}\orcidlink{0000-0002-3168-0139},
    Groom, S.~L.\inst{\ref{ipac}}\orcidlink{0000-0001-5668-3507},
    Harvey, L.\inst{\ref{dublin}}\orcidlink{0000-0003-3393-9383},
    Kasliwal, M.~M.\inst{\ref{caltechphysics}}\orcidlink{0000-0002-5619-4938},
    Kenworthy, W. D.\inst{\ref{okc}}\orcidlink{0000-0002-5153-5983},
    Kim, Y.-L.\inst{\ref{lancaster}}\orcidlink{0000-0002-1031-0796},
    Kuhn, D.\inst{\ref{lpnhe}}\orcidlink{0009-0005-8110-397X},
    Kulkarni, S.~R.\inst{\ref{caltechphysics}}\orcidlink{0000-0001-5390-8563},
    Lacroix, L.\inst{\ref{lpnhe},\ref{okc}}\orcidlink{0000-0003-0629-5746},
    Laher, R.~R.\inst{\ref{ipac}}\orcidlink{0000-0003-2451-5482},
    Masci, F.~J.\inst{\ref{ipac}}\orcidlink{0000-0002-8532-9395},
    M\"uller-Bravo, T. E.,\inst{\ref{barcelona1},\ref{barcelona2}}\orcidlink{0000-0003-3939-7167},
    Miller, A.\inst{\ref{northwestern},\ref{ciera}}\orcidlink{0000-0001-9515-478X},
    Osman, M.\inst{\ref{lpnhe}}\orcidlink{0009-0000-6101-6725},
    Perley, D.\inst{\ref{liverpool}}\orcidlink{0000-0001-8472-1996},
    Popovic, B.\inst{\ref{ip2i}}\orcidlink{0000-0002-8012-6978},
    Purdum, J.\inst{\ref{caltech}}\orcidlink{0000-0003-1227-3738},
    Qin, Y-J.\inst{\ref{caltechphysics}}\orcidlink{0000-0003-3658-6026},
    Racine, B.\inst{\ref{marseille}}\orcidlink{0000-0001-8861-3052},
    Reusch, S.\inst{\ref{desi}}\orcidlink{0000-0002-7788-628X},
    Riddle, R.\inst{\ref{caltech}}\orcidlink{0000-0002-0387-370X},
    Rosnet, P.\inst{\ref{clermont}}\orcidlink{0000-0002-6099-7565},
    Rosselli, D.\inst{\ref{marseille}}\orcidlink{0000-0001-6839-1421},
    Ruppin, F.\inst{\ref{ip2i}}\orcidlink{0000-0002-0955-8954},
    Senzel, R.\inst{\ref{dublin}}\orcidlink{0009-0002-0243-8199},
    Rusholme, B.\inst{\ref{ipac}}\orcidlink{0000-0001-7648-4142},
    Schweyer, T.\inst{\ref{okc2}}\orcidlink{0000-0001-8948-3456},
    Terwel, J. H.\inst{\ref{dublin},\ref{not}}\orcidlink{0000-0001-9834-3439},
    Townsend, A.\inst{\ref{humbolt}}\orcidlink{0000-0001-6343-3362},
    Tzanidakis, A.\inst{\ref{uw}}\orcidlink{0000-0003-0484-3331},
    Wold, A.\inst{\ref{ipac}}\orcidlink{0000-0002-9998-6732},
    Yan, L.\inst{\ref{caltech}}\orcidlink{0000-0003-1710-9339}
    }
    
   \institute{
%   Lyon
   Universite Claude Bernard Lyon 1, CNRS, IP2I Lyon / IN2P3, IMR 5822, F-69622 Villeurbanne, France \label{ip2i} 
   \and
%   Lancaster
   Department of Physics, Lancaster University, Lancs LA1 4YB, UK 
   \label{lancaster}
   \and
%   OKC
   The Oskar Klein Centre, Department of Physics, AlbaNova, SE-106 91 Stockholm , Sweden
   \label{okc}
   \and   
%   Dublin
   School of Physics, Trinity College Dublin, College Green, Dublin 2, Ireland
   \label{dublin}
   \and
%   Humbolt   
   Institut für Physik, Humboldt-Universität zu Berlin, Newtonstr. 15, 12489 Berlin, Germany
   \label{humbolt}
   \and   
%   Cambridge
    Institute of Astronomy and Kavli Institute for Cosmology, University of Cambridge, Madingley Road, Cambridge CB3 0HA, UK
    \label{cambridge}
    \and
%   OKC 2   
   The Oskar Klein Centre, Department of Astronomy, AlbaNova, SE-106 91 Stockholm , Sweden
   \label{okc2}
   \and
%   LPNHE
   Sorbonne Université, CNRS/IN2P3, LPNHE, F-75005, Paris, France
   \label{lpnhe}
   \and
%   DESI   
   Deutsches Elektronen-Synchrotron DESY, Platanenallee 6, 15738 Zeuthen, Germany
   \label{desi}
   \and
%   LBNL   
   Lawrence Berkeley National Laboratory, 1 Cyclotron Road MS 50B-4206, Berkeley, CA, 94720, USA
   \label{lbnl}
   \and
%   UCBerkeley   
    Department of Astronomy, University of California, Berkeley, 501 Campbell Hall, Berkeley, CA 94720, USA
    \label{ucb}
    \and
%   College Park
    Joint Space-Science Institute, University of Maryland, College Park, MD 20742, USA
    \label{collegepark}
    \and
    Department of Astronomy, University of Maryland, College Park, MD 20742, USA
    \label{collegepark2}
    \and
    Astrophysics Science Division, NASA Goddard Space Flight Center, Mail Code 661, Greenbelt, MD 20771, USA
    \label{goddard}
    \and    
%   Canada
    National Research Council of Canada, Herzberg Astronomy \& Astrophysics Research Centre, 5071 West Saanich Road, Victoria, BC V9E 2E7, Canada
   \label{canada}
   \and
%   Clermont   
    Université Clermont Auvergne, CNRS/IN2P3, LPCA, F-63000 Clermont-Ferrand, France
    \label{clermont} 
   \and
%   Marseille   
   Aix Marseille Université, CNRS/IN2P3, CPPM, Marseille, France
    \label{marseille}
   \and   
%   DIRAC
    DIRAC Institute, Department of Astronomy, University of Washington, 3910 15th Avenue NE, Seattle, WA 98195, USA
    \label{dirac}
    \and   
%   DUKE   
   Department of Physics, Duke University Durham, NC 27708, USA
   \label{duke}
   \and    
%   IPAC   
   IPAC, California Institute of Technology, 1200 E. California Blvd, Pasadena, CA 91125, USA
   \label{ipac}
   \and   
%   CalTech Physics    
   Division of Physics, Mathematics  \& Astronomy 249-17, Caltech. Pasadena, CA 91108, USA
   \label{caltechphysics}
   \and
%   Barcelona   
   Institute of Space Sciences (ICE-CSIC), Campus UAB, Carrer de Can Magrans, s/n, E-08193 Barcelona, Spain.
   \label{barcelona1}
   \and
   Institut d'Estudis Espacials de Catalunya (IEEC), 08860 Castelldefels (Barcelona), Spain
   \label{barcelona2}
   \and
%   Northwestern
    Department of Physics and Astronomy, Northwestern University, 2145 Sheridan Rd, Evanston, IL 60208, USA
    \label{northwestern}
    \and
%   CIERA    
    Center for Interdisciplinary Exploration and Research in Astrophysics (CIERA), Northwestern University, 1800 Sherman Ave, Evanston, IL 60201, USA
    \label{ciera}
    \and
%   Caltech   
   Caltech Optical Observatories, California Institute of Technology, Pasadena, CA 91125, USA
   \label{caltech}
   \and       
%   Liverpool    
    Astrophysics Research Institute, Liverpool John Moores University, 146 Brownlow Hill, Liverpool L3 5RF, UK
    \label{liverpool}   
   Nordic Optical Telescope, Rambla José Ana Fernández Pérez 7, ES-38711 Breña Baja, Spain
   \label{not}
   \and
%   UW   
    Department of Astronomy, University of Washington, 3910 15th Avenue NE, Seattle, WA 98195, USA
    \label{uw}
             }
   \date{}

\titlerunning{ZTF SN Ia DR2: Overview}
\authorrunning{M. Rigault et al.}
% \abstract{}{}{}{}{} 
% 5 {} token are mandatory
 
  \abstract
 {We present the first homogeneous release of several thousand
   spectroscopically classified type Ia
   supernovae (SNe~Ia) with spectroscopic redshifts.
   This release, named “DR2,” contains 3628 nearby ($z<0.3$) SNe~Ia
   discovered, followed, and classified by the \textit{Zwicky}
   Transient Facility survey between March 2018 and December 2020. Of
   these, 3000 have good-to-excellent sampling and 2667 pass standard
   cosmology light curve quality cuts. This release is thus the
   largest SN~Ia release to date, increasing by an order of magnitude
   the number of well-characterized low-redshift objects. With DR2, we
   also provide a volume-limited ($z<0.06$) sample of nearly a
   thousand SNe~Ia. With such a large, homogeneous, and
   well-controlled dataset, we are studying key current questions on
   SN cosmology, such as the linearity SNe~Ia standardization, the SN
   and host dependencies, the diversity of the SN~Ia population, and
   the accuracy of current light curve modeling. These, and more, are
   studied in detail in a series of articles associated with this
   release. 
  Alongside the SN~Ia parameters, we publish our forced-photometry
  \textit{gri}-band light curves, 5138 spectra, local and global host
  properties, observing logs, and a Python tool to facilitate the use
  and access of these data. 
  The photometric accuracy of DR2 is not yet suited for cosmological
  parameter inference, which will follow as the “DR2.5” release. We
  nonetheless demonstrate that our Hubble diagram of several thousands of
  SNe~Ia has a typical 0.15 mag scatter. 
  }


   \maketitle

\section{Introduction}

% SN Cosmology status (no tension yet)
For the last thirty years, type Ia supernovae (SNe~Ia) have played a central role in building the current standard model of cosmology. 
In the late 1990s, $\mathcal{O}(100)$ SNe~Ia led to the discovery of the acceleration of the expansion rate of the Universe, the cause of which was dubbed dark energy \citep{riess1998, perlmutter1999} (see \citet{goobar2011} for a review). 
The existence of dark energy has since been confirmed with high precision by various other cosmological probes, such as the anisotropies in the cosmic microwave background  \citep{spergel2003, planck2020} and from baryon acoustic oscillations \citep[e.g.,][]{eisenstein2005,boss2017}. 
The two decades that followed were those of the SN cosmology field maturation. Many low- ($z<0.1$) and high-redshift ($0.1<z<1$) surveys have enabled us to gather, altogether, $\mathcal{O}(1~000)$ SNe~Ia. Joining these datasets, the dark energy equation of state parameter, $w$, has been shown to be in good agreement with $w=-1$, which is expected if the acceleration of the Universe's expansion is due to a simple cosmological constant, $\Lambda$, in Einstein's general relativity equations \citep{astier2006,betoule2014,scolnic2018,brout2022}. This gain in statistics allows us to test the use of SNe~Ia as accurate standard candles of cosmology, since, despite the great success of SN cosmology, their underlying astrophysics are still largely unknown.

% The progenitor issue
It is generally accepted that an SN~Ia is the transient event resulting from the thermonuclear explosion of a carbon-oxygen white dwarf \citep{whelan73,iben1984, nugent2011}, triggered by accreting material from a companion star in a binary system \citep{liu2023}. However, the nature of the companion star (another white dwarf, main-sequence star, etc.) and the explosion mechanism (at or below the Chandrasekhar mass limit) is still unclear and no single picture has emerged \citep{maoz2014}. In recent years, sample studies of SNe Ia have been used, as they provide a population-wide approach to probe various aspects of SNe Ia physics \citep[e.g.][]{maguire2014,silverman2015,papadogiannakis2019,tucker2020,desai2024}. However, these analyses are so far limited by small-number statistics and/or survey design (e.g., coverage, cadence, and depth).
The astrophysical origin of SNe~Ia, and their homogeneity, are consequently still highly debated \citep{jha2019}. Additional data, notably spectroscopic and close to the explosion epoch, would be valuable to discriminate competitive models \citep[e.g.,][and references there in]{deckers2022}.
Nonetheless, understanding the SN~Ia mechanism is not an actual requirement for SN cosmology, as long as the astrophysical dependencies can be controlled below the statistical uncertainties. But, without a better grasp on the actual SN~Ia astrophysics, such an assertion is difficult.

%However, only a handful of SNe~Ia have exploded close enough that a direct study of the progenitor system has been possible \citep[e.g. SN\,2011fe;][]{bloom2012} and of these, no single picture have emerged \citep{maoz2014}. 

% empirical standardisation

Since the mid-90s, the success of SN cosmology has emerged from our ability to standardize their brightness from a natural scatter of $\sim0.40\,\mathrm{mag}$ down to $\sim0.15\,\mathrm{mag}$ by exploiting two empirical linear relations that correlate the SN~Ia's stretch and color, derived from their light curve, with their absolute peak brightness \citep{riess1996,tripp1998,guy2010}. A scatter of $\sim0.15\,\mathrm{mag}$ — that is, 7\% precision in distance — makes it one of the most precise standard candles in astrophysics. However, only half of this variance can be explained by measurement or light curve modeling errors.
Thus, to further control any unexplained astrophysical dependency in SN~Ia distances, the correlations between the SN properties and those of their host environments have been extensively studied over the past decade.

%  + rates (A+B) Host bias | age vs. dust | systematic dominated world

Early on, it was suggested that two populations of SNe~Ia must coexist to explain the relative rates between progenitors \citep[e.g.,][]{mannucci2005,sullivan2006,smith2012}: one population follows recent star formation (called “prompt”) with the other related to stellar mass (i.e., an old population of stars, called “delayed”). 
More delicate for cosmology, the stretch- and color-standardized SN~Ia magnitude was then shown to depend on its environment, such that those from low-mass hosts are fainter then those from high-mass galaxies \citep[e.g.,][]{sullivan2010, childress2013, roman2018}. This so called mass-step has since been used as a third standardization parameter in all cosmological analysis \citep[e.g.,][]{betoule2014,scolnic2018,brout2022, riess2022}. Yet, the origin of this bias is still highly debated. An inaccurate correction of this effect may lead to significant bias in the derivation of cosmological parameters \citep{rigault2015,rigault2020,smith2020}. Currently, the most discussed models are: differences between the host interstellar dust properties \citep{brout2021, popovic2023}, or the progenitor age, in the context of the prompt versus delayed model \citep{rigault2013,rigault2020, nicolas2021, briday2022}. Both may actually be true \citep[e.g.,][]{kelsey2021,wiseman2022}.

Today, state-of-the-art compilations of samples nearly reach 2000 SNe~Ia and are limited by systematic uncertainties \citep{brout2022,vincenzi2024}. Alongside the astrophysical biases, the dominating source of known uncertainties are calibration issues largely due to the requirement to compile SN~Ia samples from multiple surveys spanning various redshift ranges \citep[e.g.,][]{betoule2014, brout2022, vincenzi2024}. This issue is particularly critical at a low-redshift of $z<0.1$, at which dozen of samples are merged, the largest of which contains fewer than 200 targets \citep[e.g.,][]{brout2022}.

% Most recent results. H0 tension and w tension 
After two decades, the standard model of cosmology is starting to see the first hints of inconsistencies and, again, SNe~Ia are playing a key role. 
First and foremost, the direct measurement of the Hubble constant, $H_0$, derived by anchoring the absolute SN~Ia standardized luminosity with Cepheids, is found to be $5\sigma$ higher than theoretical expectations anchored by cosmic microwave background data \citep[e.g.,][]{riess2016,riess2022,planck2020}. Most recently, two independent SN~Ia compilations concluded that $w$ might differ from $w=-1$ at the 2 to $3\sigma$ level \citep{rubin2023,des2024}. These results, strengthened by the recent DESI Year-1  release \citep{desi2024}, could be the sign of new fundamental physics, or hints of unknown systematic biases in the SN~Ia distances.

% Transition
In that context, the \textit{Zwicky} Transient Facility \citep[ZTF;][]{bellm2019, graham2019} has been collecting thousands of nearby SNe a year in the northern sky since it started science operations in March 2018. As of mid-2024, $\mathcal{O}(10k)$ ZTF-discovered SNe have been spectroscopically classified. In this paper, we present an overview of the second data release associated with the “Type Ia Supernovae \& Cosmology” science working group, aka “DR2.” This dataset contains 3628 type Ia supernovae (SNe~Ia) detected, followed, and classified by the ZTF survey and discovered before the end of December 2020.

This DR2 follows \cite{dhawan2022}, in which we illustrate the characteristics of the survey using the first months of operations. Along with this release overview, 20 companion papers that study the dataset, providing key insights on nearby SN~Ia physics and their use as cosmological probes, have been released. 
However, we warn the user that the current light curve photometry calibration does not yet reach the accuracy to unlock cosmological analysis. Any other scientific study is welcomed. Ongoing work is being finished on the survey calibration that will lead to a cosmological parameter inference as the next release: “DR2.5.”

With 2667 well-sampled nearby spectroscopic SNe~Ia passing the usual light curve quality cuts, this release is the largest SN~Ia release to date, over any redshift range. It increases by an order of magnitude the current state-of-the-art low-redshift sample compilation of 192 targets used in \cite{des2024} to anchor their Hubble diagram, and by five the number of $z<0.1$  SNe~Ia ever used for cosmology \citep{scolnic2022}.

We summarize in Sect.~\ref{sec:ztf} the ZTF survey operation during the period covered by this release.  In Sect.~\ref{sec:dr2sample}, we introduce the DR2 sample, and we briefly review the release data in Sect.~\ref{sec:dr2data}. We review the 20 companion papers in Sect.~\ref{sec:dr2papers}, while the content and access of the release are summarized in Sect.~\ref{sec:dr2content}. 
We conclude in Sect.~\ref{sec:conclusion}.

\section{Summary of the ZTF operation}
\label{sec:ztf}

The ZTF survey employs the ZTF camera mounted on the P48 Schmidt
telescope at Mount Palomar Observatory. As is detailed in
\cite{dekany2020}, this  576 megapixel camera is made of sixteen
6144×6160 e2v CCD231-C6 charge-coupled devices (CCD) and equipped
with three filters: ztf:g ($g$), ztf:r ($r$), and ztf:i ($i$).  
Each CCD is subdivided into four read-out channels (called quadrants)
that have square 1.01 arcsec pixels, selected to match the typical
site image quality of $\sim$2 arcsec full width at half maximum (see
Smith et al. in prep.). 
The camera has a 47 deg$^2$ field of view with an 86.7\% filling factor. It reaches a typical 20.5 mag $5\sigma$-limit depth in 30 s, and, with a slightly less than 9 s read-out overhead made while slewing, ZTF is able to cover, at that depth, 3750 deg$^2$/hour. Between March 2018 and December 2020, ZTF acquired 480,572 exposures, with a typical same-filter cadence of three days, thanks to the National Science Foundation (NSF) -funded  Mid-Scale Innovations Program (MSIP) public survey \citep{bellm2019}. During the DR2 operations, the ZTF collaboration also operated a number of additional surveys, including an extragalactic high-cadence survey, with six visits a night in the \textit{g} and \textit{r} bands, a larger-area ($\sim1800$ deg$^2$) survey that acquired same-night exposures, mostly in $g$ and $r$, and an \textit{i}-band survey with a four-day cadence \citep{bellm2019b}.

ZTF also has a low-resolution integral field spectrograph
\citep[SEDm,][]{blagorodnova2018,rigault2019,kim2022, lezmy2022}
dedicated to spectroscopically classifying transients detected by the
photometric survey. The Bright Transient Survey (BTS) ZTF cross
working group is designed to use the SEDm, and other spectrographs, to
construct a magnitude-limited spectroscopically classified SN
sample. As is detailed in \cite{perley2020}, they reach a 97\%, 93\%,
and 75\% completeness for objects brighter than 18 mag, $18.5$ mag,
and $19$ mag, respectively \citep[see also][]{fremling2020}. The vast
majority of our targets (79\%) are covered by the BTS program. The
remaining targets are usually fainter than 19 mag (see Smith et al. in
prep.).

\section{3628 type Ia supernovae}
\label{sec:dr2sample}

This DR2 consists of SNe Ia observed by ZTF between March 2018 and December 2020, with the date cutoff based on the discovery date of the SN. The composition of this sample is detailed in Table~\ref{tab:sample} and reviewed below.

% ======================= %
%      SUMMARY TABLE      %
% ======================= %
\begin{table}
\centering
\tiny
\caption{ZTF cosmology science working group DR2 sample statistics}
\label{tab:sample}
\begin{tabular}{l c c c}
\hline\\[-0.8em]
\hline\\[-0.5em]
Cuts                & n targets & removed & \% removed  \\[0.15em]
\hline\\[-0.5em]
Master list         & 3795  & --    & --    \\[0.15em]
+ ZTF light curve    & 3778  & 17    & 0.4   \\[0.30em]
+ a spectrum        & 3668  & 110   & 2.9   \\[0.15em]
+ confirmed “Ia”    & \textbf{3628} & 40    & 1.1   \\[0.30em]

\hline\\[-0.5em]
Basic cuts\\[0.30em]
\hline\\[-0.5em]
Good light curve sampling  & 2960  & 668   & 18.4  \\[0.30em]
$x_1 \in [-3,+3]$   & 2899  & 61    & 2.1    \\[0.30em]
$c \in [-0.2,0.8]$  & 2861  & 38    & 1.4    \\[0.30em]
$\sigma_{t_0}\leq 1$  & 2836  & 25    & 0.9    \\[0.30em]
$\sigma_{x_1}\leq 1$  & 2822  & 14    & 0.5    \\[0.30em]
$\sigma_{c}\leq 0.1$  & 2809  & 13    & 0.4    \\[0.30em]
“fitprob”$ \geq 10^{-7}$  & \textbf{2667}  & 142    & 5.1    \\[0.30em]
% 2849 (+7% if usual field good sampling cut.)
\hline\\[-0.5em]
Subsample examples\\[0.30em]
\hline\\[-0.5em]
%good "fitprob" & 2445 & 464 & 16.0 \\[0.30em]
Volume limited ($z\leq0.06$) & 994  & 1673  & 62.7  \\[0.30em]
Non-peculiar SNe~Ia & 2629  & 38   & 1.4 \\[0.30em]
\hline\\[-0.5em]
\end{tabular}
\tablefoot{“Subsample examples” are “logical or” cuts starting from the 2667 SNe~Ia.}
\end{table}

To build this dataset, we started from the list of any target that has been flagged, at some point, as an “SN~Ia” either from our internal databases \citep{kasliwal2019, duev2019, vdwalt2019, coughlin2023}, or through the Transient Name Server.\footnote{\url{www.wis-tns.org}} This “master list” contains 3795 targets and, of these, 17 have been removed, since they have no ZTF light curves but happened to be in our databases.

Since we aim to provide spectroscopically confirmed SNe~Ia, we request to be able to release a spectrum that leads to a classification for each target. 
Therefore, 110 objects flagged as “SN~Ia” without a spectrum, or based on non-publicly available spectra, have been discarded from this release. This corresponds to 2.9\% of the initial targets with ZTF light~curves. Of the remaining 3668 SNe with at least one spectrum, there were 40 for which a secure Ia classification was not possible. We are thus releasing data for 3628 spectroscopically confirmed SNe Ia.

Following a careful study of the light curve fit residuals, 
we concluded in \cite{dr2rigault} that the rest-frame phase range,
$\phi \in [-10,+40]$, is sufficiently trained to derive reasonable light curve model fits.
Therefore, considering this phase-range only,
we defined targets with “good sampling” as those that have, at least:  detections at seven phases, two of which are before and two of which are after peak luminosity ($\phi\equiv0$), and detections in two photometric bands. Here, we refer to as “phases” the rest-frame phase with respect to the estimated maximum light ($t_0$; see Sect.~\ref{sec:dr2data_dist}), while same-night same-band detections are ignored, such that, for example, four detections on the same night in $g$ only account for 1 phase. These strict phase coverage criteria reduce the number of targets by 18.4\%, leaving 2960 SNe Ia.

\begin{figure}
  \centering
  \includegraphics[width=\linewidth]{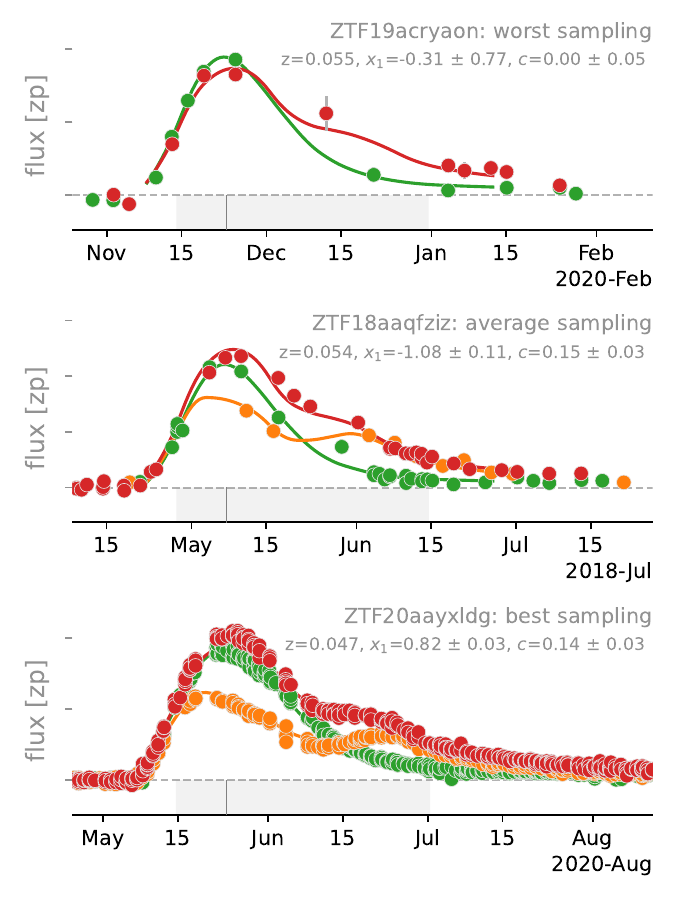}
  \caption{Three SN Ia light curve examples illustrating, from top to bottom, the worst, average, and best sampling of DR2. Photometric points observed with the $g$, $r$, and $i$ bands are shown as green, red, and orange markers, respectively. Lines show the best fit SALT2 model with matching colors, and the associated parameters in the legend. The lower gray bands show the $\phi\in[-10, +40]$ rest-frame phase range used to fit the SALT model. The vertical lines indicate the estimated maximum light.
  }
  \label{fig:lcexamples}
\end{figure}

The norm of the SN cosmology fields usually relaxes the “two pre-max
detection” criteria to “a detection prior to $\phi=+5$ days”
\citep[e.g.,][]{betoule2014, scolnic2022}. Doing so would leave 3244
objects. However, good sampling pre- and post-maximum is preferred to
ensure the correct estimation of the light curve parameters.  We
illustrate in Fig.~\ref{fig:lcexamples} an example of an SN Ia with
among the worst light curve sampling (seven phases), along with SNe Ia
with average (40 phases) and best sampled (130 phases) light curves
(see light curve coverage statistics in Smith et al. in prep).

We used the SALT2 light curve fitter \citep{guy2010, betoule2014, taylor2021} to fit our light curves. To ensure the resulting parameters are reasonable for use in cosmological measurements, we further implemented the following criteria (see Table~\ref{tab:sample}): $-3\leq x_1 \leq3$; $-0.2\leq c\leq0.8$; $\sigma_{x_1}<1$ ; $\sigma_{c}<0.1$; $\sigma_{t_0}<1$; a quality of fit (“fitprob”) $\geq10^{-7}$; with $\sigma_x$ the measured errors on the $x$ SALT2 parameters and with fitprob derived from the best SALT fit $\chi^2$  \citep[see e.g.][]{scolnic2018}.
This left 2667 SNe~Ia that pass all our cosmology-ready quality cuts. Using the aforementioned looser light curve coverage criteria that has been used in previous cosmology studies, we would have 2849 targets instead.

Of these 2667 SNe Ia, 994 have a redshift lower than $z=0.06$, up to
which our sample of non-peculiar SNe~Ia is free from nonrandom selection functions, aka a “volume-limited sample” (see
\citealt{dr2amenouche} and detailed study of this sample in
\citealt{dr2ginolina} and \citealt{dr2ginolinb}): up to
  $z<0.06$ the observed distribution of any parameter (e.g., stretch,
  color, peak-magnitude) is representative of the parameter's
  underlying distribution. Hence, non-peculiar ZTF DR2 SNe~Ia up to
  $z=0.06$ are free from Malmquist bias, and the parameter
  distributions and their correlations should be representative of
  that given by nature (see dedicated discussion on
  Sect.~\ref{sec:dr2data_selection}).

Careful subclassification indicates that 2625 of these 2667 SNe Ia are suitable for SN cosmology; that is, non-peculiar objects. This involves removing peculiar subclasses like “91bg-like” and ``Ia-csm” objects, but keeping subclasses generally used in cosmological studies, like ``91T-like'' or ``99aa-like'' SNe Ia (for further details of these subclassifications and how they were determined, see \cite{dr2dimitriadis} and \cite{dr2burgaza}).

\section{Overview of the released data}
\label{sec:dr2data}

Details concerning the DR2 data and the derived parameters are given
in Smith et al. (in prep.). 
This section summarizes the top-level information.

\subsection{Light curves}
\label{sec:dr2data_lc}

In this release, we provide $g$, $r$, and $i$ ZTF light curves for all
our 3628 SNe Ia acquired by the ZTF camera installed on the
P48. Additional photometric data points, such as those obtained from the SED
machine in camera mode, are not included in this release. 

\subsubsection{sampling statistics}
\label{sec:samplingstats}
On the 2960 SNe Ia passing the good sampling criteria, only 9 do not have $g$-band and only 2 do not have $r$-band detections in the rest-frame phase range of $-$10 to +40 with respect to maximum light. However, only 46\% of our targets have $i$-band detections in this same phase range. Most targets acquired since early 2021 will have coverage in all three bands as we updated our observing strategy, but they are not part of this sample. Future releases of DR2 targets will include additional $i$-band observations, after the remaining reference images are processed. For these 2960 targets, we typically have 40 detections (median) in the $\phi\in[-10,+40]$ day phase range, with medians of 9 and 27 points pre- and post-maximum light, respectively. This goes up to more than 166 detections for the top 10\% most observed sample that is part of the aforementioned high-cadence regions. On average, our SNe~Ia are detected 12 d before maximum light and up to 55 d after, with a median one-day cadence in any filter. This corresponds to a typical 2.9, 2.5, and 6 d same-filter revisit for the $g$, $r$, and $i$ bands, respectively. Figure~\ref{fig:lcexamples} illustrates the typical worst, average, and best sampling light curve of the release, with the increased uncertainty in the derived SALT2 values of $x_1$ and $c$ with decreasing sampling, shown in the legend. 

\subsubsection{Photometry}
\label{sec:photometry}

As is detailed in Smith et al. (in prep.), the ZTF SN Ia DR2 light curves have been extracted using a custom recalibration of the forced-photometry pipeline presented in \cite{masci2019} and \cite{yao2019}. 
The absolute zero-point, set to $zp=30$ in the released light curves,
is known at the 5 percent level, but the relative photometry is closer
to 1\%. Our limited knowledge of the photometry stems from the use of
forced photometry, since reference stars are not measurable in
difference imaging. For the next ZTF SN Ia releases, we shall use
scene-modeling photometry \citep[e.g.,][]{holtzman2008, astier2013,
  brout2019} that do not require difference images to extract the
transient light~curve, and thus 
enable one to use the same flux estimator on both stars and the
transients of interest. 
We demonstrate in Lacroix et al. (in prep.) that we have a working
scene-modeling  pipeline, and by comparing it to the released DR2
light~curves we can assert that our light~curve colors (i.e., relative
calibration) are good at the percent level (see details in Smith
  et al. in prep.). 
However, Lacroix et al. (in prep.) identified a nonlinear bias in the
flux measurements, on the percent level, dubbed the “pocket effect.”
This effect was observed starting in November of 2019, when the camera
wave-front read-out system characteristics had been updated. The
origin of this effect, its modeling, and its correction are presented
in Lacroix et al. (in prep.), but the current ZTF light~curves are affected by this read-out issue.

\begin{figure}
  \centering
  \includegraphics[width=\linewidth]{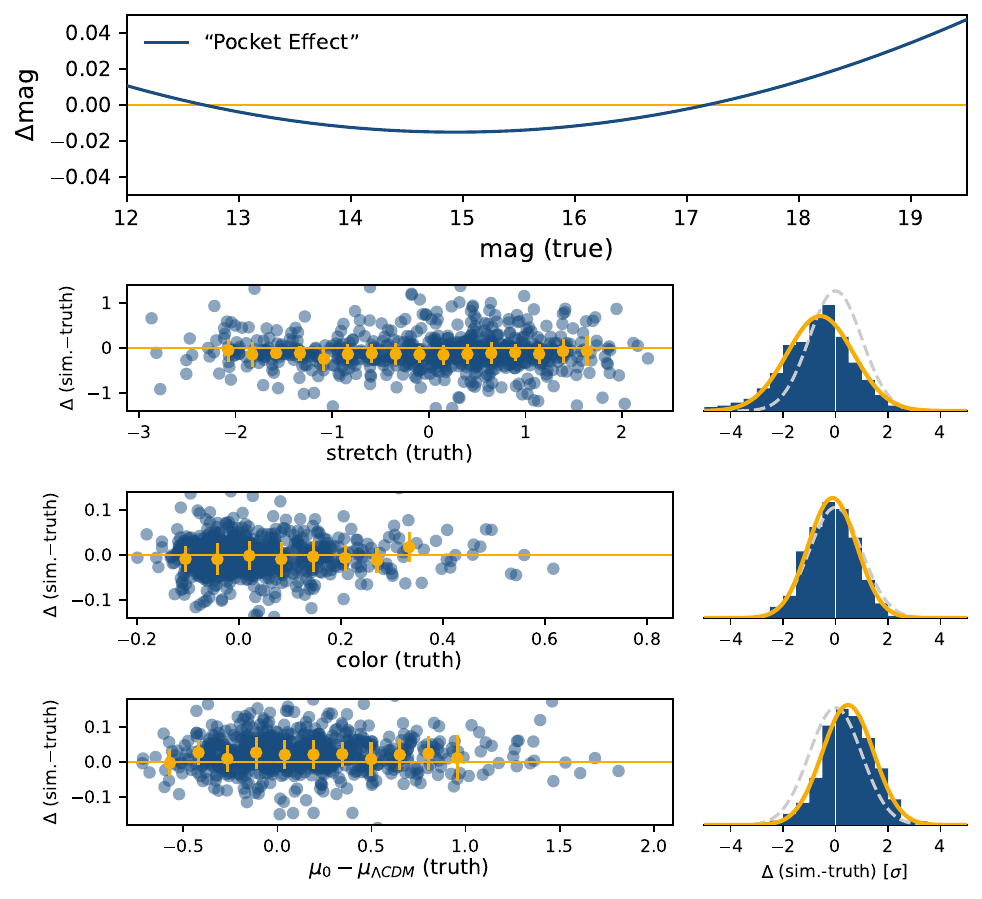}
  \caption{Impact of the pocket effect photometric nonlinearity affecting ZTF data since November 2019.
  \emph{Top panel}: Magnitude bias caused by the pocket effect (model) as a function of (input) magnitude.
  \emph{Lower panels}: Difference between SN~Ia parameters recovered from realistic light~curve simulations affected by the pocket effect as a function of the input parameter (truth). \emph{Top to bottom}: SN Ia peak magnitude in $b$ band (from $x_0$), the SN Ia stretch, color, and Hubble residual (non-standardized). 
  Orange markers (errors) show the median (nmad) difference per bin of input parameters. The horizontal orange lines show zero.
  The histograms on the right display parameter pulls ((sim.-truth) or error), which should follow a $\mathcal{N}(0,1)$ distribution (dashed gray). Best fit normal distributions on pulls are shown in orange. In the $m_b$ panel, the top gray ticks show the redshift corresponding to the peak magnitude of a typical $M_B=-19.3$ SN Ia.
  }
  \label{fig:pocketeffect}
\end{figure}

We illustrate in Fig.~\ref{fig:pocketeffect} (top) the typical impact of the pocket effect on our data: in comparison to their true magnitudes, they deviate by up to a few percent peak to peak. 
To test the impact of this effect on our results, we simulated realistic light~curves, as in \cite[][see also our Sect.~\ref{sec:dr2data_selection}]{dr2amenouche}, and we perturbed them with the nonlinearity effect presented in the top panel of Fig.~\ref{fig:pocketeffect}. We then derived their light~curve parameters, as is presented in Sect.~\ref{sec:dr2data_dist}. 
The resulting shifts are shown in the bottom panels of Fig.~\ref{fig:pocketeffect}.
The pocket effect is expected to exacerbate fainter SNe Ia and bias their observed stretch standardization parameter. As was expected, the light~curve colors are not affected by the sensor issue. 
However, the stretch recovered is typically $\Delta\,x_1=-0.14$ (70\% of the average stretch error, $\sigma_{x_1}$) lower than what is given as an input. This average shift is independent of the actual input value and could be seen as a simple stretch zero-point definition bias. 
As is discussed by \cite{dr2ginolina}, who study the stretch standardization, this stretch zero-point bias has no significant impact on our results, since the absolute stretch value is meaningless as long as we do not compare the ZTF DR2 stretch values with these from another survey, as was done, for instance, in \cite{nicolas2021}. In such a case, we suggest applying a $\Delta\,x_1=-0.14$ correction term.

The SN Ia peak magnitude slowly drifts because of the pocket effect, especially beyond  18.5 mag, as is shown in the $m_b$ panel of Fig.~\ref{fig:pocketeffect}; so typically SN~Ia at $z>0.08$. But when focusing on the Hubble residuals, a similar trend to that of the stretch is observed: the zero point is off by $\sim0.02\,\mathrm{mag}$ ($0.5\sigma$, when not accounting for the intrinsic scatter) and this shift is uncorrelated with the actual Hubble residuals value. Consequently, the pocket effect has no impact on pure ZTF DR2 magnitude residual studies (e.g., steps), since this zero-point cancels out with cosmology ($M_0$ definition).

We emphasize that only targets acquired after November 2019 are affected by the pocket effect. Hence, we encourage people using ZTF SN~Ia DR2 data to test their results when only including the 50\% of the ZTF SN~Ia DR2 sample with a peak magnitude prior to October 2019 as was done, for example, in \cite{dr2ginolina}.
However, since the ZTF SN~Ia DR2 light~curves are derived from forced photometry and the abovementioned pocket effect, we discourage the use of this data release for precision cosmological parameter inference. 
The next release (DR2.5, planned for the end of 2025), will address
both problems, and hence will be suitable for precision cosmology (see
Lacroix et al. in prep.). While it will remove a systematic bias, 
we nonetheless do not expect any significant reduction in the Hubble residuals dispersion once the pocket effect is corrected, given its negligible amplitude in comparison to the SN~Ia intrinsic scatter.

%# Spectra
\subsection{Spectra}
\label{sec:dr2data_spectra}

This DR2 contains 5138 spectra associated with the 3628 SNe Ia in our spectroscopically confirmed sample. 
Each target has at least one spectrum with sufficient quality to enable a secured “Ia” classification, and 28\% have multiple spectra. Most of our spectra (60\%) have been acquired by the SEDm and, thanks to the observing strategy of the BTS program, the vast majority of these have been acquired near or before maximum light. In addition to the SEDm, we obtained data from many other facilities, such as the Liverpool Telescope \citep[7.6\%;][]{steele2004}, the Palomar 200-inch Hale Telescope \citep[7.3\%;][]{}, and European Southern Observatory's New Technology Telescope (NTT), as part of the ePESSTO program \citep[6.5\%;][]{smartt2015}.

All of the spectra were matched with the \texttt{snid} classification algorithm \citep{blondin2007} using a custom template library made of 370 templates (available upon request). This matching was used as an initial classification indicator and to derive SN-feature-based redshifts for all our targets (for further details of how further subclassifications were performed see Sect. \ref{sec:dr2data_subclass}, as well as \cite{dr2dimitriadis} and \cite{dr2burgaza}).

%# Host
\subsection{Host galaxies}
\label{sec:dr2data_host}

As is detailed in Smith et al. (in prep.), the host identification of each
SN Ia in our sample was made in two steps. First, we queried public
databases, such as  DESI-LS DR9 \citep[][]{dey2019}, SDSS DR17
\cite[][]{Abdurrouf2022}, and PS1 DR2 \citep[][]{flewelling2020}, for
sources within an 100~kpc radius, given the estimated redshifts from \textsc{snid} spectral template matching. 
Then, we computed the directional light radius
\citep[DLR,][]{sullivan2006,gupta2016} between the SN and each
surrounding galaxy or galaxy-like source and identified the host as
the source with the smaller DLR. The SNe for which no galaxy has a
$DLR<7$ were excluded.  
The global photometry was derived using the \texttt{HostPhot} package \citep{muller2022} on public $g$, $r$, $i$, $z$, $y$ PS1 DR2 images that cover the same sky as ZTF. 
Local photometry was made using the 2~kpc radius aperture photometry of these data \citep[e.g.][]{briday2022}. Once optical global and local photometry had been estimated, we computed stellar masses and rest-frame color using a spectral energy distribution fitting performed with the \texttt{P{\`E}GASE2} galaxy spectral templates \citep{leborgne2002}, as in, for example, \cite{sullivan2010}.

%# Typing
\subsection{(Sub)-Classification}
\label{sec:dr2data_subclass}

From the initial master list of 3668 targets that have at least a ZTF light curve and a spectrum, we collectively visually inspected each target to ensure they were indeed an SN Ia and to subclassify them when possible. This process was made in two steps. Firstly, the ZTF Cosmology SWG members manually inspected the data through a specially designed web application.\footnote{typingapp.in2p3.fr} In the end, 32 users did more than 14,000 individual (sub)-classifications, which fed a decision-tree algorithm to automatically (sub)-classify the objects. This way, each SN~Ia from the DR2 was vetted at least twice, and 3.5 times on average. 

In a second step, SN~Ia population experts double-checked edge cases (including those for which ambiguous subtypes were suggested) in the volume-limited ($z<0.06$) sample to release the final (sub)-classification (see details in \citealt{dr2dimitriadis} and \citealt{dr2burgaza}). At the end of this procedure, about 5\% of the SN~Ia targets were classified as too peculiar to be included in cosmological analyses (e.g., ``Ia-CSM'' or ``91bg-like''). However, some unidentified peculiar subtypes likely remain in the sample at $z>0.06$ that were not checked in detail. These contaminants are likely limited to overluminous events, such as ``Ia-CSM'' and ``03fg-like'' SNe Ia that would be detected at these higher redshifts. These events are intrinsically rare and would also likely fail the SALT2 light curve quality cuts that are applied for cosmological measurements. 

%# redshift
\subsection{Redshift}
\label{sec:dr2data_redshift}

Each of our 3628 targets has a redshift that comes from one of four sources in order of preference: the public galaxy redshift catalog (2200, 60.6\%), galaxy emission lines visible in non-SEDm target spectra (199, 5.5\%), galaxy emission lines visible in SEDm spectra (121, 3.3\%), and estimated using \textsc{snid} SN~Ia  template matching (1086, 30.0\%). The first two cases have a typical precision on the redshift, $\sigma_z \leq 10^{-4}$, and are referred to as ``galaxy-redshifts,'' while the latter have a precision of $\sigma_z \approx 10^{-3}$ and are referred to as ``SN redshifts.'' Of the 2200 galaxy catalog redshifts, 71\% come from the DESI MOST Hosts program \citep{soumagnac2024}.

\begin{figure}
  \centering
  \includegraphics[width=\linewidth]{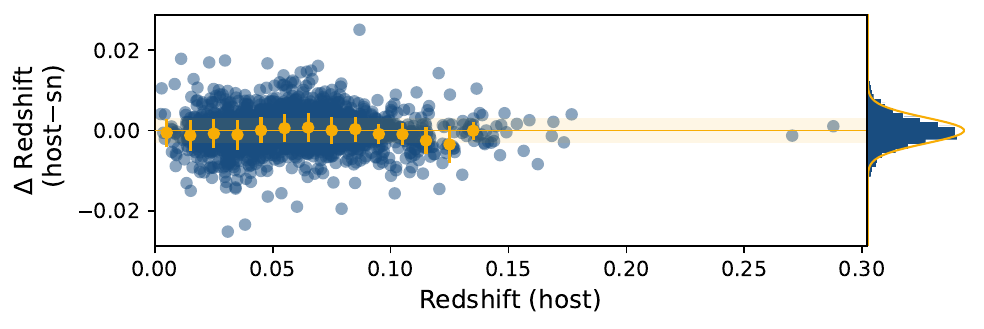}
  \caption{Comparison between the host redshift (mostly from DESI) and SN-feature-based redshifts. Orange markers represent the median and nmad (error bar) per bins of $\delta z=0.01$ redshift. 
  The horizontal orange line shows “zero,” while the orange band shows the  $\pm3\times10^{-3}$ range. The right-panel histogram shows the $\Delta z$ distribution overplotted (orange) with a normal distribution, $\mathcal{N}(0, 3\times10^{-3})$.
  }
  \label{fig:snidz}
\end{figure}

Figure~\ref{fig:snidz} illustrates the accuracy and precision of the
SN-feature based redshifts that account for 30\% of the sample (22\%
for the volume-limited sample, $z<0.06$) using the SN-feature
redshifts of SNe~Ia with galaxy redshifts \citep[mostly DESI,
see][]{soumagnac2024}. We conclude that SN-feature redshifts are
unbiased (average shifts lower than $10^{-5}$) with a typical scatter
of $3\,10^{-3}$ across the entire redshift range coverage by ZTF (see
details in Smith et al. in prep).
Such a $\sigma_z \approx 3\,10^{-3}$ scatter ($900\,\mathrm{km\ s^{-1}}$) typically corresponds to a 0.09 mag additional scatter on the Hubble diagram at our median redshift of $z_\mathrm{med}=0.07$.

Redshift are given in the heliocentric frame and the DR2 redshift distribution is shown in Fig.~\ref{fig:scatterx1cz}. Since most of our hosts are brighter than 20 mag in the $r$ band, the vast majority of our SN redshifts will soon be acquired and released by DESI as part of their bright galaxy survey program \citep{hahn2023}.

\subsection{Selection}
\label{sec:dr2data_selection}

In Sect.~\ref{sec:dr2data}, we claimed that our sample is volume-limited up to $z=0.06$. A fully realistic simulation based on observing log analysis supporting this claim is presented in \cite{dr2amenouche}. In this section, we provide key elements characterizing the sample selection of our dataset.

\begin{figure}
  \centering
  \includegraphics[width=\linewidth]{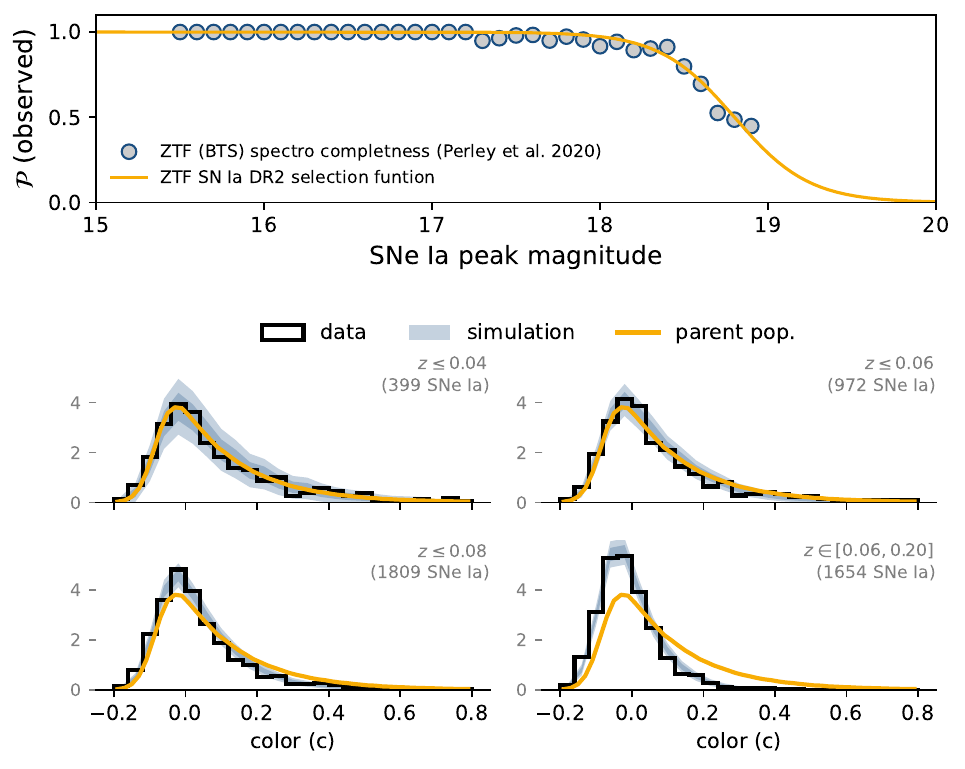}
  \caption{
  Selection function of the ZTF SN Ia DR2 sample and its effect on the color distribution. 
  \emph{Top}: Probability that a SN Ia spectrum is acquired and leads to a classification as a function of SN Ia peak magnitude. The model is the survival sigmoid function $1-\mathcal{S}(18.8, 4.5)$.
  \emph{Bottom}: Distribution of the ZTF SN Ia DR2 color (black) compared to the expected parent population (orange) and simulation predictions ($1$ and $2\sigma$, light blue) for various redshift ranges (see legend).
  }
  
  \label{fig:selection}
\end{figure}

As is presented in Smith et al. in prep., the ZTF survey photometric depth is 20.4 mag in $g$, 20.6 mag in $r$ and 20.0 mag in $i$ on average, but the spectroscopic follow-up is $\sim1.5$ mag shallower. As we require a spectroscopic classification for the ZTF SN~Ia DR2 release, this spectroscopic follow-up magnitude limit  consequently is our limiting selection function.
Figure~\ref{fig:selection} (top) presents our effective selection function, modeled as a survival sigmoid, $1-\mathcal{S}(m\,;\,m_0, s)= 1-\left( 1+e^{s(m-m_0)}  \right)^{-1}$, with $m_0=18.8$ and $s=4.5$. As was expected, this model closely matches the BTS spectroscopic typing completeness measurements from \cite{perley2020}.

The steep fainter-redder SN Ia relation means that the first
  SNe~Ia missed by a magnitude-limited survey are red targets (see
  e.g., discussion in \citealt{nicolas2021}). Hence, the SN~Ia color
  distribution is highly sensitive to such a selection, more so than,
  for instance, the SN~Ia stretch. We show in Fig.~\ref{fig:selection}
  the color distribution of non-peculiar SNe~Ia of the DR2 sample for
  various redshift ranges ($z\leq0.045$, $z\leq0.06$, $z\leq0.08$, and
  $z\in[0.06, 0.20]$). We overplot our parent population model,
  finding it to be within $1-2\sigma$, using a scatter derived from
  3000 realistic simulations of the same number of targets than that
  observed in the DR2. For the parent population, we assume an
  intrinsic Gaussian distribution convolved with an extrinsic
  exponential decay given by \cite{dr2ginolinb} (see also e.g., \cite{brout2021} and \cite{dr2popovic}). 

We draw two conclusions from this figure. 
First, since the simulations, which are sensitive to both the assumed parent population and the modeled selection function, closely match the observed data for all of the various redshift ranges, we conclude that our model for the selection and underlying SNIa distributions are reasonable. 
Second, since the observed SN~Ia DR2 color distribution at $z<0.06$ closely matches the parent population (significant deviations are visible in the $z<0.08$ case), and since the color distribution is the most sensitive to selection effects, we conclude that our sample is not affected by a significant selection function up to $z\leq0.06$; that is, it is volume-limited up to $z=0.06$ (see additional details in \cite{dr2amenouche}).

\subsection{Distances}
\label{sec:dr2data_dist}

We fit our light curves using the SALT2 \citep{guy2007,guy2010} algorithm in its version 4 \citep{betoule2014} retrained by \cite{taylor2021} and made available by the \texttt{sncosmo}  package\footnote{https://sncosmo.readthedocs.io/en/stable/about.html} as “SALT2-T21.” 
As is detailed in Smith et al. in prep, the fit was performed in the $\phi\in[-10, +40]$ d rest-frame phase range as this is when the light curve algorithm is sufficiently trained (see also \citealt{dr2rigault}). To do so, the fit was performed twice. First, given a $t_0$ guess coming from the light curve data, we cut at $\phi\in[-15, +50]$ d and fit SALT2 to get a robust $t_0$, then we applied the $\phi\in[-10, +40]$ d cut to refit the light curve and store the SALT2 parameters, their errors, and covariances.

\begin{figure}
  \centering
  \includegraphics[width=1\linewidth]{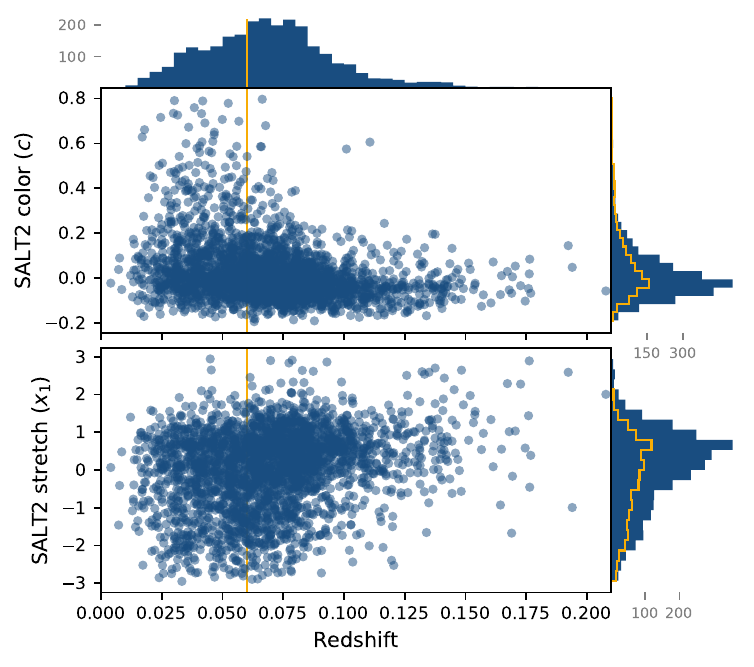}
  \caption{Distribution of SALT2 stretch ($x_1$) and color ($c$) parameters as a function of redshift for the 2667 SNe~Ia passing the basic quality cuts detailed in Table~\ref{tab:sample}. 
    The vertical yellow lines illustrate the redshift below which the DR2 sample can be considered volume-limited ($z<0.06$) for normal SNe Ia. The corresponding volume-limited parameter distributions are shown in yellow in the marginalized histograms.} 
  \label{fig:scatterx1cz}
\end{figure}

We show the SALT2 stretch ($x_1$) and color ($c$) parameters for all the 2667 SNe Ia passing our “basic quality” cuts in Fig~\ref{fig:scatterx1cz}.
Above $z=0.06$, it can be seen that the fraction of red (high $c$) and slow (low $x_1$) SNe~Ia starts to decrease rapidly due to selection effects, since these faster and redder SNe~Ia are fainter. Below this redshift of 0.06, our sample is considered to be free from nonrandom selection functions for the normal SN~Ia population (see Sect.~\ref{sec:dr2data_selection}). We refer to this sample as “volume limited” and it contains nearly 1000 normal SN Ia targets.

\begin{figure*}
  \centering
  \sidecaption
  \includegraphics[width=12cm]{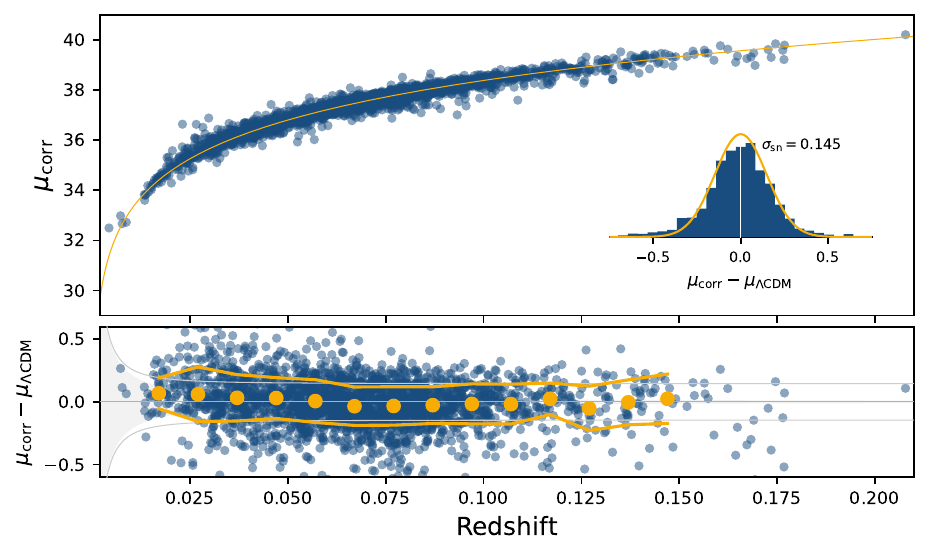}
  \caption{ZTF SN Ia DR2 Hubble diagram.
  \emph{Top:} Hubble diagram of the 2629 standardized non-peculiar SNe Ia compared to the $\Lambda$CDM model anchored on \citep{planck2020} with an arbitrary zero-point offset.
  \emph{Bottom} Standardized Hubble residuals. Orange markers (lines) show the median ($\pm$nmad) data per bin of redshifts ($\Delta z=0.01$). The light gray band shows the expected scatter caused by peculiar velocities of 300 km s$^{-1}$, while gray lines present the expected total scatter, adding in quadrature a 0.145 mag scatter (see text).
  \emph{Inset panel:} Standardized Hubble residual distribution (blue) with a normal distribution centered on zero with a scatter of    0.145 mag, adding in quadrature a velocity dispersion (0.149 mag).
  }
  \label{fig:hubblediagram}
\end{figure*}

We present our ZTF Hubble diagram for the 2629 non-peculiar SNe~Ia that pass our quality cuts in Fig.~\ref{fig:hubblediagram}. These SNe Ia have been standardized using the methodology detailed in \cite{dr2ginolina}, given the SALT2 stretch and color parameter and host local environmental properties \citep[e.g.,][]{sullivan2010, rigault2020}. Following that paper, we have used $\alpha=-0.16$, $\beta=3.05$, and $\gamma=0.145\,\mathrm{mag}$ (local-color step).

Before standardization, our SNe~Ia have a natural scatter along the Hubble diagram of 0.33~mag, using the normalized median absolute deviation (nMAD) as a robust scatter estimator. The standard deviation (std) is 0.46 mag (nMAD=std for a normal distribution).
This scatter reduces down to 0.165~mag after standardization (std: 0.209 mag), accounting for all 2629 of the ZTF DR2 non-peculiar SNe~Ia.

Part of this scatter is due to peculiar motions that typically are $\mathcal{O}(300)$ km s$^{-1}$, leading to an additional scatter of 0.03 mag at our median redshift ($z_\mathrm{med}=0.07$), but to more than 0.1~mag for targets with $z\leq0.02$, as is illustrated in Fig.~\ref{fig:hubblediagram}. 
Another part is caused by the use of SN-feature redshifts for 32\% of the sample. As is presented in Sect.~\ref{sec:dr2data_redshift}, these have a dispersion of $\sim$900 km s$^{-1}$, corresponding to an additional Hubble residual scatter of $0.09$ mag at $z_\mathrm{med}=0.07$. 
To estimate our Hubble scatter without the impact of peculiar velocities and SN-feature redshifts, we took the scatter in the range of $z\in[0.03, 0.1]$. There, the standardized SN~Ia scatter is 0.150~mag (nMAD), corresponding to 0.145~mag after removing the expected peculiar velocity contribution. This $\sigma_{SN}=0.145~\mathrm{mag}$ scatter is illustrated in the inset panel of Fig.~\ref{fig:hubblediagram}, showing it is a good description of our 2629 SN~Ia standardized Hubble residual dispersion. Such a 0.15 magnitude scatter is in good agreement with the state-of-the-art results from the literature \citep[see e.g.,][]{brout2022}.

We finally inspected for outliers that may affect our standard deviation (std) measurements (nMAD statistics have been robust). Given our sample size, Chevaunet's criteria (a less than 50\% chance of detecting such an outlier) correspond to a $4.1\,\sigma$ rejection. Doing so removed 23 objects (0.8\% of the total sample) and led to a total standard deviation (std) of $\sigma_{SN}=0.187\,\mathrm{mag}$ (nMAD: 0.159 mag), including 2606 SN~Ia, and $\sigma_{SN}=0.166\,\mathrm{mag}$ (nMAD: 0.144 mag) when discarding SN-feature redshift SNe~Ia (1647 left, $3.95\sigma$ rejection).

% ======================= %
%      Papers             %
% ======================= %

\section{Data release papers}
 \label{sec:dr2papers}

 This data release contains 3628 spectroscopically confirmed SNe~Ia. 
 In this Letter, we present a high-level overview of these data, but readers are referred to the 20 companion papers (Table~\ref{tab:dr2papers}) for more in-depth information on the sample, as well as analyses covering sample calibration, SN Ia physics, and cosmological applications. In this section, we summarize the topics covered in these DR2 papers.  

Data acquisition and processing (light curve extraction, SALT2
fitting, host matching, redshifts, etc.) are presented in Smith et
al. in prep., while Lacroix et al. in prep. review the accuracy of the
DR2 photometry and ongoing work to unlock cosmology-ready
calibrations.  An overview of the DR2 SN~Ia spectra can be found in
Johansson et al. in prep.

%Modelling 
The accuracy of the light curve modeling is reviewed in \cite{dr2rigault}.  An additional investigation of light curve modeling is presented in \cite{dr2kenworthy}, in which we introduce the possibility of an extra stretch-like light curve parameter that may absorb a significant part of the usual phase-independent color term. 

%simulations and standardisation
The simulation from \cite{dr2amenouche} and data distributions
presented in Smith et al. in prep. and in this overview show that our normal SNe~Ia dataset is free from nonrandom selections at $z<0.06$. Based on this volume-limited sample, \cite{dr2ginolina} have analysed the SN~Ia standardization process and demonstrated the nonlinearity of the “bright-slower” relation, while highlighting the most significant environmental magnitude offset to date. The origin of this offset and the color standardization are further discussed in \cite{dr2ginolinb} and compared with other higher-redshift samples in \cite{dr2popovic}. The SN~Ia standardization is studied through siblings in \cite{dr2dhawan}.

%Environments and cosmology
Given our large sample statistics, we compared the SN~Ia properties as a function of their cosmic web origin. \cite{dr2ruppin} study how SNe~Ia vary as a function of their cluster association, notably since cluster galaxies are less star-forming than their field counterparts. \cite{dr2aubert} generalize this study when comparing SN~Ia properties as a function of their cosmic density field. The actual impact of peculiar motion caused by the velocity field in the derivation of cosmological parameters is discussed in \cite{dr2carreres}.

%Progenitors 1 - overviews
 Our sample also enables careful studies of SN~Ia explosion physics and progenitor origins. \cite{dr2burgaza} present the spectroscopic diversity (through measurements of key spectral features) of the maximum-light volume-limited sample. This includes an investigation of the impact of host-galaxy contamination on spectral measurements, as well as an analysis, including spectral modeling, of the continuum of SN Ia spectral properties seen from the normal SN Ia population to peculiar underluminous subclasses, such as 91bg-like events. \cite{dr2dimitriadis} review the photometric diversity of the SN~Ia population of the DR2 (including peculiar events), describing how subclassifications have been derived, as well as intrinsic rates of the subclasses.

 %Progenitors II - specifics & hosts
In \cite{dr2terwel}, we search for late-time interaction with circumstellar material in SNe Ia, finding a few cases of rebrightening years after the explosion. In \cite{dr2harvey}, we study the demographic of high-velocity silicon features, finding that most SNe~Ia exhibit such behavior at early phases and that they are more common in underluminous SNe Ia. \cite{dr2deckers} use Gaussian process fitting of the ZTF light curves to study the properties of the second maximum clearly visible in redder bands, as well as constrain the origin of these features. The connection between host and SNe~Ia properties is further studied in \cite{dr2burgaza}, who investigate the low-mass host SN~Ia population, and in \cite{dr2senzel}, who compare the SN~Ia properties and whether they originate from the bulge, bar, or disk of the host that is connected to the age and metallicity.

%Cosmology to come
A cosmological analysis will soon follow, once photometric
calibrations detailed in Lacroix et al. in prep. are completed. This resulting DR2.5 release will be accompanied by a series of calibration papers. We discourage the user from using the current DR2 data to derive cosmological parameters.

\begin{table}
\centering
\tiny
\caption{Overview of the ZTF DR2 paper release.}
\label{tab:dr2papers}
\begin{tabular}{l c}
\hline\\[-0.8em]
\hline\\[-0.5em]
First Author   & Short title   \\[0.15em]
\hline\\[-0.5em]
% France

Rigault (a, this work)   &   DR2 overview                \\[0.30em]
Smith       &   DR2 data review             \\[0.30em]
Lacroix     &   DR2 photometry              \\[0.30em]
Johansson   &   DR2 spectra review          \\[0.30em]
Rigault (b) &   Light curve residuals        \\[0.30em]
Kenworthy   &   Light curve modeling         \\[0.30em]
Amenouche   &   DR2 sample simulations             \\[0.30em]
Ginolin (a) &   Host, stretch \& steps      \\[0.30em]
Ginolin (b) &   Host, color \& bias origin  \\[0.30em]
Popovic     &   Host \& color evolution     \\[0.30em]
Dhawan      &   SNe~Ia siblings             \\[0.30em] 
Ruppin      &   SNe~Ia in clusters          \\[0.30em] 
Aubert      &   SNe~Ia in voids             \\[0.30em] 
Carreres    &   Velocity systematics        \\[0.30em]
Burgaz (a)  &   SN~Ia spectral diversity    \\[0.30em]
Dimitriadis &   Thermonuclear SN diversity\\[0.30em]
Terwel      &   Late-time CSM interaction        \\[0.30em]
Harvey      &   High-velocity features      \\[0.30em]
Deckers     &   Secondary maxima           \\[0.30em]
Burgaz (b)  &   SNe~Ia in low-mass hosts        \\[0.30em]
Senzel      &   Bulge vs. Disk SNe~Ia       \\[0.30em]
\hline\\[-0.5em]
\end{tabular}
\end{table}

% ======================= %
%       Science           %
% ======================= %

\section{Data release content and access}
\label{sec:dr2content}

\begin{figure}
  \centering
  \includegraphics[width=0.9\linewidth]{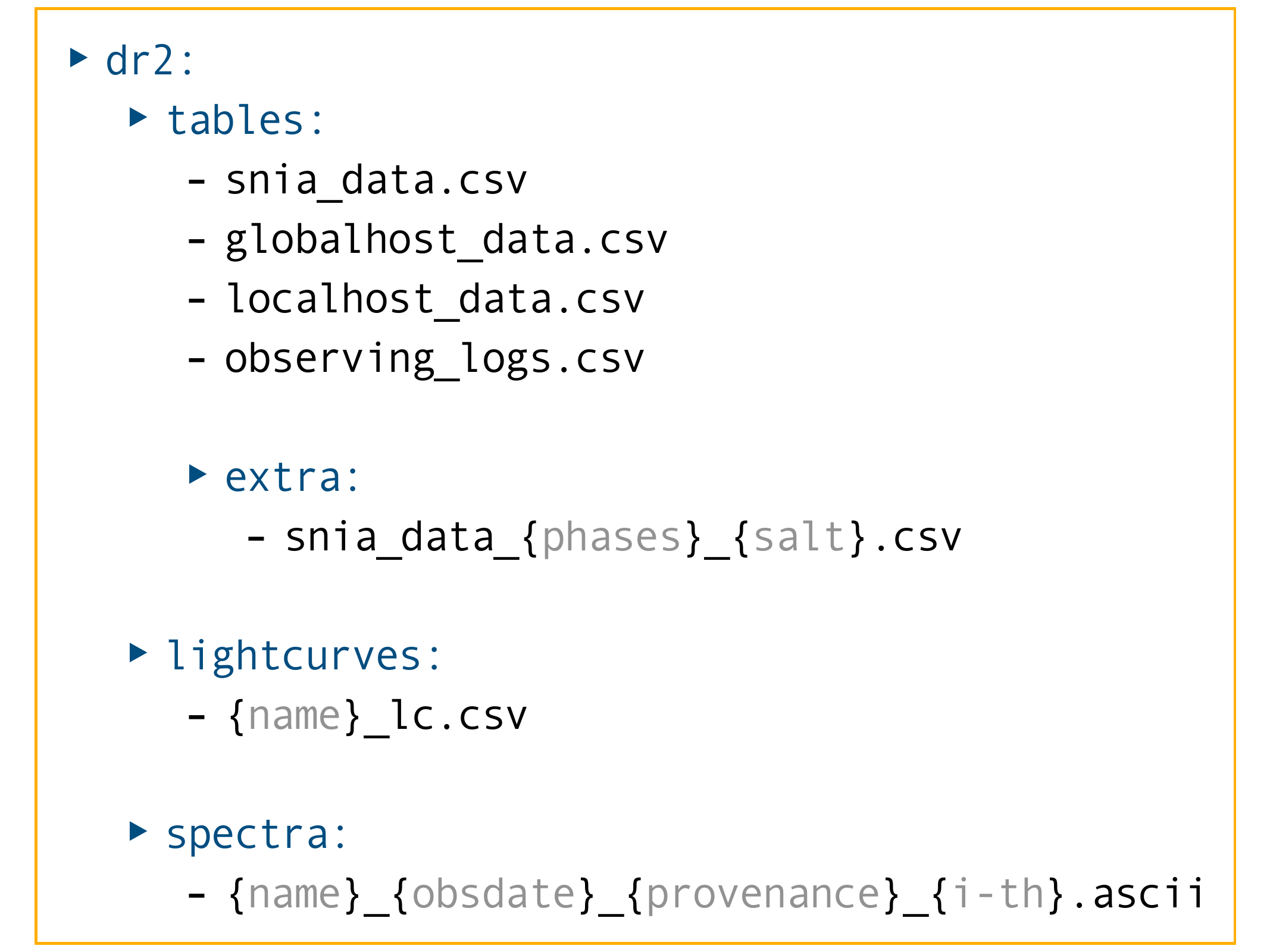}
  \caption{Content of the released data repository. The blue items are
    directories and \{\texttt{gray}\}  entries represent multiple
    items. The \texttt{observing\_logs} table cannot be automatically
    downloaded (too large, see text). 
  }
  \label{fig:repocontent}
\end{figure}

The content of this data release is illustrated in Fig.~\ref{fig:repocontent} and the main release parameters are summarized in Table~\ref{tab:dr2params}. We are providing:
\begin{itemize}
    \item 3591 SNe~Ia light curves ($g$, $r$, and $i$ band), 
    \item 5138 spectra, with at least one per SN~Ia,
    \item an SN metadata table,
    \item two host data tables (global and local properties),
    \item observing logs.
\end{itemize}

\begin{table}
\centering
\tiny
\caption{Main release table parameters.}
\label{tab:dr2params}
\begin{tabular}{l  c}
\hline\\[-0.8em]
\hline\\[-0.5em]
column   &  comment   \\[0.15em]
\hline\\[-0.5em]
\multicolumn{2}{l}{Light curve file parameters}\\[0.30em]
\hline\\[-0.5em]
mjd & observation modified julian date [day]\\[0.30em]
filter & used filter (ztfg, ztfr, ztfi)\\[0.30em]
flux & blind flux in unit of zero-point near 30\\[0.30em]
flux\_err & flux error\\[0.30em]
flag** & bit-mask (bad: $[1, 2, 4, 8, 16]$)\\[0.30em]
field\_id & ZTF sky field ID\\[0.30em]
rcid & ZTF camera quadrant ID\\[0.30em]
\hline\\[-0.5em]
\multicolumn{2}{l}{SN metadata table}\\[0.30em]
\hline\\[-0.5em]
redshift  & heliocentric     \\[0.30em]
redshift\_err  & do not incl. method scatter     \\[0.30em]
source & redshift estimation method\\[0.30em]
t0* & modified julian date [day]\\[0.30em]
x0* & blinded flux zeropoint near 30\\[0.30em]
x1* & SALT2 stretch\\[0.30em]
c* & SALT2 color\\[0.30em]
mwebv & Assumed milky way dust with $R_V=3.1$\\[0.30em]
fitprob & --\\[0.30em]
ra, dec & SN~Ia coordinates [deg]\\[0.30em]
sn\_type & SN~Ia classification\\[0.30em]
sub\_type & subclassification if any.\\[0.30em]
lccoverage\_flag & passes the good sampling cut (bool, Table~\ref{tab:sample})\\[0.30em]
fitquality\_flag & passes all other Basic cuts (bool, Table~\ref{tab:sample})\\[0.30em]
\hline\\[-0.5em]
\multicolumn{2}{l}{Global Host properties}\\[0.30em]
\hline\\[-0.5em]
ra, dec & host coordinates [deg]\\[0.30em]
d$_\mathrm{DLR}$ & normalized direct light distance\\[0.30em]
mass & (log) stellar mass (from SED fit)\\[0.30em]
color & k-corrected $g-z$ color (mag)\\[0.30em]
\hline\\[-0.5em]
\multicolumn{2}{l}{Local (2~kpc radius) environmental properties}\\[0.30em]
\hline\\[-0.5em]
mass & (log) stellar mass (from SED fit)\\[0.30em]
color & k-corrected $g-z$ color (mag)\\[0.30em]
\hline\\[-0.5em]
\multicolumn{2}{l}{Observing logs}\\[0.30em]
\hline\\[-0.5em]
mjd & modified Julian date [day]\\[0.30em]
filter & used filter (ztfg, ztfr, ztfi)\\[0.30em]
fieldid & ztf grid field index\\[0.30em]
ra, dec & ztf central footprint coordinates\\[0.30em]
rcid & amplifier index (1->64)\\[0.30em]
maglim & limited 5$\sigma$ (point source)\\[0.30em]
gain & amplifier gain\\[0.30em]
expid & exposure id.\\[0.30em]
\hline\\[-0.5em]
\end{tabular}
\tablefoot{*: also contain \texttt{\_err} for error values and \texttt{cov\_$a$\_$b$} for covariance between $a$ and $b$ terms.
**: See details in Smith et al. in prep.).}
\end{table}

We have not released light curve data and derived light curve properties for the 37 non-peculiar SNe~Ia below $z<0.015$ ($\sim65\,\mathrm{Mpc}$) from which independent distances could be acquired and that will lead to a measurement of the Hubble constant, $H_0$. These SNe~Ia will be released as part of a dedicated ZTF $H_0$ cosmology program. However, light curve data of the 12 peculiar SNe Ia at $z<0.015$ SNe~Ia have been released, as well as spectra and host properties for all targets, including the normal very nearby SNe~Ia.

The host tables contains the rest-frame $g-z$ color and stellar mass
($\log(M_*/M_\odot)$) both for local (2~kpc radius aperture) and
global properties. The host table also contains the host coordinates
and SN-host distance information. The log table contains pointing
(mjd, ra, dec) and observing conditions (zp, limiting magnitude, gain,
infobits), which is sufficient to simulate ZTF DR2 data. 

We finally provide additional SN~Ia data tables in the \texttt{tables/extra} repository. These are results obtained when changing the used SALT template (SALT2.4 \citealt{betoule2014} or SALT3 \citealt{kenworthy2021}, see modeling discussion in \cite{dr2rigault}), or changing the considered phase range (starting at $-20,\ -15,\ -10$ or $-5$ d, and/or finishing at $+30,\ +40,\ +45,\ +50$ days) or the use or not of the $i$ band.

All of the released data can be retrieved from
\href{http://ztfcosmo.in2p3.fr}{ztfcosmo.in2p3.fr}.\footnote{Data
  tables available after publication} Our observations are also made
available via WISeREP \citep{yarongalyam2012}.\footnote{https://www.wiserep.org}

\section{Conclusions}
\label{sec:conclusion}

With 3628 objects, we are releasing the largest sample of SN Ia data to date. This release, named DR2, publishes all SNe Ia acquired by the ZTF survey up to December 2020 and spectroscopically classified as “Ia.” The classification as well as the data content have been vetted by multiple members of the scientific collaboration. Along with 20 companion papers studying in detail SN~Ia astrophysics and their use as cosmological distance indicators, we are releasing: 3591 forced-photometry light curves ($g$, $r$, $i$) and SALT2 light curve parameters, 5138 spectra, galaxy and SN redshifts, global and local (2~kpc) host properties, and observing logs. The SN light curves and light curve properties of normal SNe~Ia at $z<0.015$, however, have not been released.

This release increases by an order of magnitude the current state-of-the-art low-redshift SN~Ia dataset. Our sample is homogeneous, with well-controlled selection effects. As part of our release, nearly 1000 SNe~Ia are from a volume-limited sample ($z<0.06$) for which parameter distributions and correlations are representative of the true, normal underlying SN~Ia population. 

The photometric accuracy of the release is at the percent level with known percent nonlinearity. The data are thus of sufficient quality for any internal ZTF SN~Ia analysis but are not yet ready to be joined with other datasets to derive cosmological parameters. 
The Cosmology Science Working group is actively working on final calibration steps, to release the cosmology associated with these SNe~Ia. This will constitute the DR2.5 release that will become public on a timescale of approximately one year. The next step will be the use of the full ZTF survey (2018-2025), which should contain nearly 8,000 spectroscopically confirmed SNe~Ia. In addition, we expect 25,000 additional photometrically confirmed SNe~Ia. This $\mathcal{O}(30~000)$ SNe~Ia sample will be our DR3.

\begin{acknowledgements}
Based on observations obtained with the Samuel Oschin Telescope 48-inch and the 60-inch Telescope at the Palomar Observatory as part of the \textit{Zwicky} Transient Facility project. ZTF is supported by the National Science Foundation under Grants No. AST-1440341 and AST-2034437 and a collaboration including current partners Caltech, IPAC, the Weizmann Institute of Science, the Oskar Klein Center at Stockholm University, the University of Maryland, Deutsches Elektronen-Synchrotron and Humboldt University, the TANGO Consortium of Taiwan, the University of Wisconsin at Milwaukee, Trinity College Dublin, Lawrence Livermore National Laboratories, IN2P3, University of Warwick, Ruhr University Bochum, Northwestern University and former partners the University of Washington, Los Alamos National Laboratories, and Lawrence Berkeley National Laboratories. Operations are conducted by COO, IPAC, and UW.
SED Machine is based upon work supported by the National Science Foundation under Grant No. 1106171
The ZTF forced-photometry service was funded under the Heising-Simons Foundation grant \#12540303 (PI: Graham).
The Gordon and Betty Moore Foundation, through both the Data-Driven Investigator Program and a dedicated grant, provided critical funding for SkyPortal.
This project has received funding from the European Research Council (ERC) under the European Union's Horizon 2020 research and innovation program (grant agreement n 759194 - USNAC). This project is supported by the H2020 European Research Council grant no. 758638. This work has been supported by the Agence Nationale de la Recherche of the French government through the program ANR-21-CE31-0016-03.
L.G. acknowledges financial support from the Spanish Ministerio de Ciencia e Innovaci\'on (MCIN) and the Agencia Estatal de Investigaci\'on (AEI) 10.13039/501100011033 under the PID2020-115253GA-I00 HOSTFLOWS project, from Centro Superior de Investigaciones Cient\'ificas (CSIC) under the PIE project 20215AT016 and the program Unidad de Excelencia Mar\'ia de Maeztu CEX2020-001058-M, and from the Departament de Recerca i Universitats de la Generalitat de Catalunya through the 2021-SGR-01270 grant. Y.-L.K. has received funding from the Science and Technology Facilities Council [grant number ST/V000713/1]. This work has been enabled by support from the research project grant ‘Understanding the Dynamic Universe’ funded by the Knut and Alice Wallenberg Foundation under Dnr KAW 2018.0067
This work has been supported by the research project grant “Understanding the Dynamic Universe” funded by the Knut and Alice Wallenberg Foundation under Dnr KAW 2018.0067,  {\em Vetenskapsr\aa det}, the Swedish Research Council, project 2020-03444, and the G.R.E.A.T research environment, project number 2016-06012.
SRK’s research is supported by the Heising-Simons Foundation. ZTF is supported in part by the Medium Scale Instrumentation Program (MSIP of the National Science Foundation (NSF)
LH is funded by the Irish Research Council under grant number GOIPG/2020/1387.
ECB acknowledges support by the NSF AAG grant 1812779 and grant \#2018-0908 from the Heising-Simons Foundation. MMK acknowledges generous support from the David and Lucille Packard Foundation.

\end{acknowledgements}


\begin{thebibliography}{}
% A
\bibitem[Abdurro'uf et al.(2022)]{Abdurrouf2022} Abdurro'uf, Accetta, K., Aerts, C., et al.\ 2022, \apjs, 259, 35.

\bibitem[Amenouche et al.(2024)]{dr2amenouche} Amenouche, M., Smith,
  M., Rosnet, P., et al.\ 2024,
  arXiv:2409.04650. doi:10.48550/arXiv.2409.04650, (ZTFSI)
  
\bibitem[Astier et al.(2006)]{astier2006} Astier, P., Guy, J., Regnault, N., et al.\ 2006, \aap, 447, 31.

\bibitem[Astier et al.(2013)]{astier2013} Astier, P., El Hage, P., Guy, J., et al.\ 2013, \aap, 557, A55.

\bibitem[Aubert et al.(2024)]{dr2aubert} Aubert, M.,  Rosnet, P., Popovic, B., et al.\ 2024, arXiv:2406.11680. doi:10.48550/arXiv.2406.11680, (ZTFSI)

% B
\bibitem[Bellm et al.(2019)]{bellm2019} Bellm, E.~C., Kulkarni, S.~R., Graham, M.~J., et al.\ 2019, \pasp, 131, 018002.

\bibitem[Bellm et al.(2019)b]{bellm2019b} Bellm, E.~C., Kulkarni, S.~R., Barlow, T., et al.\ 2019, \pasp, 131, 068003.

\bibitem[Betoule et al.(2014)]{betoule2014} Betoule, M., Kessler, R., Guy, J., et al.\ 2014, \aap, 568, A22.

\bibitem[Blagorodnova et al.(2018)]{blagorodnova2018} Blagorodnova, N., Neill, J.~D., Walters, R., et al.\ 2018, \pasp, 130, 035003.

\bibitem[Bloom et al.(2012)]{bloom2012} Bloom, J.~S., Kasen, D., Shen, K.~J., et al.\ 2012, \apjl, 744, L17.

\bibitem[Blondin \& Tonry(2007)]{blondin2007} Blondin, S. \& Tonry, J.~L.\ 2007, \apj, 666, 1024.

\bibitem[BOSS Collab.(2017)]{boss2017} Alam, S., Ata, M., Bailey, S., et al.\ 2017, \mnras, 470, 2617.

\bibitem[Briday et al.(2022)]{briday2022} Briday, M., Rigault, M., Graziani, R., et al.\ 2022, \aap, 657, A22.

\bibitem[Brout et al.(2019)]{brout2019} Brout, D., Sako, M., Scolnic, D., et al.\ 2019, \apj, 874, 106.

\bibitem[Brout \& Scolnic(2021)]{brout2021} Brout, D. \& Scolnic, D.\ 2021, \apj, 909, 26.

\bibitem[Brout et al.(2022)]{brout2022} Brout, D., Scolnic, D., Popovic, B., et al.\ 2022, \apj, 938, 110.

\bibitem[Burgaz et al.(2024,a)]{dr2burgaza} Burgaz, U., Maguire, K.,
  Dimitriadis, G., et al.\ 2024, arXiv:2407.06828. (ZTFSI)

 \bibitem[Burgaz et al.(2024,b)]{dr2burgaza} Burgaz, U., Maguire, K., Burgaz, U. et al.\ 2024, \aap, submitteb (ZTFSI)
  
%  C
\bibitem[Carreres et al.(2024)]{dr2carreres} Carreres, B.,
  Rosselli, D., Bautista, J.~E., et al.\ 2024,
  arXiv:2405.20409. doi:10.48550/arXiv.2405.20409, (ZTFSI)
  
\bibitem[Childress et al.(2013)]{childress2013} Childress, M., Aldering, G., Antilogus, P., et al.\ 2013, \apj, 770, 108.

\bibitem[Coughlin et al.(2023)]{coughlin2023} Coughlin, M.~W., Bloom, J.~S., Nir, G., et al.\ 2023, \apjs, 267, 31.

% D
\bibitem[Deckers et al.(2022)]{deckers2022} Deckers, M., Maguire, K., Magee, M.~R., et al.\ 2022, \mnras, 512, 1317.

\bibitem[Deckers et al.(2024)]{dr2deckers} Deckers, M., Maguire, K., Shingles, L., et al.\ 2024, arXiv:2406.19460. doi:10.48550/arXiv.2406.19460, (ZTFSI)
  
\bibitem[Dekany et al.(2020)]{dekany2020} Dekany, R., Smith, R.~M., Riddle, R., et al.\ 2020, \pasp, 132, 038001.

\bibitem[Desai et al.(2024)]{desai2024} Desai, D.~D., Kochanek, C.~S., Shappee, B.~J., et al.\ 2024, \mnras, 530, 5016.
  
\bibitem[DES Collab.(2024)]{des2024} DES collaboration, Abbott, T.~M.~C., Acevedo, M., et al.\ 2024, arXiv:2401.02929. doi:10.48550/arXiv.2401.02929

\bibitem[DESI Collab.(2024)]{desi2024} DESI Collaboration, Adame, A.~G., Aguilar, J., et al.\ 2024, arXiv:2404.03002. doi:10.48550/arXiv.2404.03002

\bibitem[Dey et al.(2019)]{dey2019} Dey, A., Schlegel, D.~J., Lang, D., et al.\ 2019, \aj, 157, 168.

\bibitem[Dhawan et al.(2022)]{dhawan2022} Dhawan, S., Goobar, A., Smith, M., et al.\ 2022, \mnras, 510, 2228.

\bibitem[Dhawan et al.(2024)]{dr2dhawan} Dhawan, S.,
  Mortsell, E., Johansson, J., et al.\ 2024,
  arXiv:2406.01434. doi:10.48550/arXiv.2406.01434, (ZTFSI)
  
\bibitem[Dimitriadis et al.(2024)]{dr2dimitriadis} Dimitriadis, G.,
  Burgaz, U., Deckers, M., et al.\ 2024,
  arXiv:2409.04200. doi:10.48550/arXiv.2409.04200, (ZTFSI)
  
\bibitem[Duev et al.(2019)]{duev2019} Duev, D.~A., Mahabal, A., Masci, F.~J., et al.\ 2019, \mnras, 489, 3582.

% E
\bibitem[Eisenstein et al.(2005)]{eisenstein2005} Eisenstein, D.~J., Zehavi, I., Hogg, D.~W., et al.\ 2005, \apj, 633, 560.
% F
\bibitem[Flewelling et al.(2020)]{flewelling2020} Flewelling, H.~A., Magnier, E.~A., Chambers, K.~C., et al.\ 2020, \apjs, 251, 7.

\bibitem[Fremling et al.(2020)]{fremling2020} Fremling, C., Miller, A.~A., Sharma, Y., et al.\ 2020, \apj, 895, 32.

% G
\bibitem[Ginolin et al.(2024,a)]{dr2ginolina} Ginolin, M., Rigault, M., Smith, M., et al.\ 2024, arXiv:2405.20965. doi:10.48550/arXiv.2405.20965, (ZTFSI)

\bibitem[Ginolin et al.(2024,b)]{dr2ginolinb} Ginolin, M., Rigault, M., Copin, Y., et al.\ 2024, arXiv:2406.02072. doi:10.48550/arXiv.2406.02072, (ZTFSI)

\bibitem[Goobar \& Leibundgut(2011)]{goobar2011} Goobar, A. \& Leibundgut, B.\ 2011, Annual Review of Nuclear and Particle Science, 61, 251.

\bibitem[Graham et al.(2019)]{graham2019} Graham, M.~J., Kulkarni, S.~R., Bellm, E.~C., et al.\ 2019, \pasp, 131, 078001.

\bibitem[Gupta et al.(2016)]{gupta2016} Gupta, R.~R., Kuhlmann, S., Kovacs, E., et al.\ 2016, \aj, 152, 154.

\bibitem[Guy et al.(2007)]{guy2007} Guy, J., Astier, P., Baumont, S., et al.\ 2007, \aap, 466, 11.

\bibitem[Guy et al.(2010)]{guy2010} Guy, J., Sullivan, M., Conley, A., et al.\ 2010, \aap, 523, A7.

% H
\bibitem[Hahn et al.(2023)]{hahn2023} Hahn, C., Wilson, M.~J., Ruiz-Macias, O., et al.\ 2023, \aj, 165, 253.

\bibitem[Harvey et al.(2024)]{dr2harvey} Harvey, L., Maguire, K., Burgaz, U. et al.\ 2024, \aap, submitteb (ZTFSI)
  
\bibitem[Holtzman et al.(2008)]{holtzman2008} Holtzman, J.~A., Marriner, J., Kessler, R., et al.\ 2008, \aj, 136, 2306.

% I
\bibitem[Iben \& Tutukov(1984)]{iben1984} Iben, I. \& Tutukov, A.~V.\ 1984, \apjs, 54, 335. doi:10.1086/190932

% J
\bibitem[Jha et al.(2019)]{jha2019} Jha, S.~W., Maguire, K., \& Sullivan, M.\ 2019, Nature Astronomy, 3, 706.

% K
\bibitem[Kasliwal et al.(2019)]{kasliwal2019} Kasliwal, M.~M., Cannella, C., Bagdasaryan, A., et al.\ 2019, \pasp, 131, 038003.

\bibitem[Kelsey et al.(2021)]{kelsey2021} Kelsey, L., Sullivan, M., Smith, M., et al.\ 2021, \mnras, 501, 4861.

\bibitem[Kenworthy et al.(2021)]{kenworthy2021} Kenworthy, W.~D., Jones, D.~O., Dai, M., et al.\ 2021, \apj, 923, 265.
\bibitem[Kenworthy et al.(2024)]{dr2kenworthy} Kenworthy, W.~D., Goobar, A., Jones, D.~O. et al.\ 2024, \aap, submitteb (ZTFSI)
  
\bibitem[Kim et al.(2022)]{kim2022} Kim, Y.-L., Rigault, M., Neill, J.~D., et al.\ 2022, \pasp, 134, 024505.

% L
\bibitem[Le Borgne \& Rocca-Volmerange(2002)]{leborgne2002} Le Borgne, D. \& Rocca-Volmerange, B.\ 2002, \aap, 386, 446.

\bibitem[Lezmy et al.(2022)]{lezmy2022} Lezmy, J., Copin, Y., Rigault, M., et al.\ 2022, \aap, 668, A43. 

\bibitem[Liu et al.(2023)]{liu2023} Liu, Z.-W., R{\"o}pke, F.~K., \& Han, Z.\ 2023, Research in Astronomy and Astrophysics, 23, 082001. 

% M
\bibitem[Maguire et al.(2014)]{maguire2014} Maguire, K., Sullivan, M., Pan, Y.-C., et al.\ 2014, \mnras, 444, 3258.

\bibitem[Mannucci et al.(2005)]{mannucci2005} Mannucci, F., Della Valle, M., Panagia, N., et al.\ 2005, \aap, 433, 807.

\bibitem[Maoz et al.(2014)]{maoz2014} Maoz, D., Mannucci, F., \& Nelemans, G.\ 2014, \araa, 52, 107.

\bibitem[Masci et al.(2019)]{masci2019} Masci, F.~J., Laher, R.~R., Rusholme, B., et al.\ 2019, \pasp, 131, 018003.

\bibitem[M{\"u}ller-Bravo \& Galbany(2022)]{muller2022} M{\"u}ller-Bravo, T. \& Galbany, L.\ 2022, JOSS, 7, 4508.

% N
\bibitem[Nicolas et al.(2021)]{nicolas2021} Nicolas, N., Rigault, M., Copin, Y., et al.\ 2021, \aap, 649, A74. 

\bibitem[Nugent et al.(2011)]{nugent2011} Nugent, P. E., Sullivan, M., Cenko, S. B., et al.\ 2011, Nature, 480, 344.
% O
  
% P
\bibitem[Papadogiannakis et al.(2019)]{papadogiannakis2019} Papadogiannakis, S., Dhawan, S., Morosin, R., et al.\ 2019, \mnras, 485, 2343. 

\bibitem[Perley et al.(2020)]{perley2020} Perley, D.~A., Fremling, C., Sollerman, J., et al.\ 2020, \apj, 904, 35. 

\bibitem[Perlmutter et al.(1999)]{perlmutter1999} Perlmutter, S., Aldering, G., Goldhaber, G., et al.\ 1999, \apj, 517, 565. 

\bibitem[Planck Collab.(2020)]{planck2020} Planck Collaboration, Aghanim, N., Akrami, Y., et al.\ 2020, \aap, 641, A6. 

\bibitem[Popovic et al.(2023)]{popovic2023} Popovic, B., Brout, D.,
  Kessler, R., et al.\ 2023, \apj, 945,
  84. doi:10.3847/1538-4357/aca273
  
\bibitem[Popovic et al.(2024)]{dr2popovic} Popovic, B.,
  Rigault, M., Smith, M., et al.\ 2024,
  arXiv:2406.06215. doi:10.48550/arXiv.2406.06215, (ZTFSI)

% Q

% R
\bibitem[Roman et al.(2018)]{roman2018} Roman, M., Hardin, D., Betoule, M., et al.\ 2018, \aap, 615, A68

\bibitem[Riess et al.(1996)]{riess1996} Riess, A.~G., Press, W.~H., \& Kirshner, R.~P.\ 1996, \apj, 473, 88.

\bibitem[Riess et al.(1998)]{riess1998} Riess, A.~G., Filippenko, A.~V., Challis, P., et al.\ 1998, \aj, 116, 1009. 

\bibitem[Riess et al.(2016)]{riess2016} Riess, A.~G., Macri, L.~M., Hoffmann, S.~L., et al.\ 2016, \apj, 826, 56. 

\bibitem[Riess et al.(2022)]{riess2022} Riess, A.~G., Yuan, W., Macri, L.~M., et al.\ 2022, \apjl, 934, L7. 

\bibitem[Rigault et al.(2013)]{rigault2013} Rigault, M., Copin, Y., Aldering, G., et al.\ 2013, \aap, 560, A66.

\bibitem[Rigault et al.(2015)]{rigault2015} Rigault, M., Aldering, G., Kowalski, M., et al.\ 2015, \apj, 802, 20.

\bibitem[Rigault et al.(2019)]{rigault2019} Rigault, M., Neill, J.~D., Blagorodnova, N., et al.\ 2019, \aap, 627, A115.

\bibitem[Rigault et al.(2020)]{rigault2020} Rigault, M., Brinnel, V., Aldering, G., et al.\ 2020, \aap, 644, A176.

  
  \bibitem[Rigault et al.(2024,b)]{dr2rigault} Rigault, M.,
    Smith, M., Regnault, N., et al.\ 2024,
    arXiv:2406.02073. doi:10.48550/arXiv.2406.02073, (ZTFSI)
    
\bibitem[Rubin et al.(2023)]{rubin2023} Rubin, D., Aldering, G.,
  Betoule, M., et al.\ 2023,
  arXiv:2311.12098. doi:10.48550/arXiv.2311.12098
  
  \bibitem[Ruppin et al.(2024)]{dr2ruppin} Ruppin, F.,
    Rigault, M., Ginolin, M., et al.\ 2024,
    arXiv:2406.01108. doi:10.48550/arXiv.2406.01108, (ZTFSI)

    % S
\bibitem[Senzel et al.(2024)]{dr2senzel} Senzel, R., Maguire, K.,
  Burgaz, U., et al.\ 2024,
  arXiv:2411.11986. doi:10.48550/arXiv.2411.11986 (ZTFSI)

\bibitem[Scolnic et al.(2018)]{scolnic2018} Scolnic, D.~M., Jones, D.~O., Rest, A., et al.\ 2018, \apj, 859, 101.

\bibitem[Scolnic et al.(2022)]{scolnic2022} Scolnic, D., Brout, D., Carr, A., et al.\ 2022, \apj, 938, 113.

\bibitem[Silverman et al.(2015)]{silverman2015} Silverman, J.~M., Vink{\'o}, J., Marion, G.~H., et al.\ 2015, \mnras, 451, 1973.

\bibitem[Smartt et al.(2015)]{smartt2015} Smartt, S.~J., Valenti, S., Fraser, M., et al.\ 2015, \aap, 579, A40. 

\bibitem[Smith et al.(2012)]{smith2012} Smith, M., Nichol, R.~C., Dilday, B., et al.\ 2012, \apj, 755, 61. 

\bibitem[Smith et al.(2020)]{smith2020} Smith, M., Sullivan, M., Wiseman, P., et al.\ 2020, \mnras, 494, 4426. 

\bibitem[Soumagnac et al.(2024)]{soumagnac2024} Soumagnac, M.~T., Nugent, P., Knop, R.~A., et al.\ 2024, arXiv:2405.03857. doi:10.48550/arXiv.2405.03857

\bibitem[Spergel et al.(2003)]{spergel2003} Spergel, D.~N., Verde, L., Peiris, H.~V., et al.\ 2003, \apjs, 148, 175. 

\bibitem[Steele et al.(2004)]{steele2004} Steele, I.~A., Smith, R.~J., Rees, P.~C., et al.\ 2004, \procspie, 5489, 679. 

\bibitem[Sullivan et al.(2006)]{sullivan2006} Sullivan, M., Le Borgne, D., Pritchet, C.~J., et al.\ 2006, \apj, 648, 868.

\bibitem[Sullivan et al.(2010)]{sullivan2010} Sullivan, M., Conley, A., Howell, D.~A., et al.\ 2010, \mnras, 406, 782. 

% T
\bibitem[Taylor et al.(2021)]{taylor2021} Taylor, G., Lidman, C., Tucker, B.~E., et al.\ 2021, \mnras, 504, 4111. 

\bibitem[Terwel et al.(2024)]{dr2terwel} Terwel, J.~H., Maguire, K.,
  Dimitriadis, G., et al.\ 2024,
  arXiv:2402.16962. doi:10.48550/arXiv.2402.16962, (ZTF SI)

\bibitem[Tripp(1998)]{tripp1998} Tripp, R.\ 1998, \aap, 331, 815

\bibitem[Tucker et al.(2020)]{tucker2020} Tucker, M.~A., Shappee, B.~J., Vallely, P.~J., et al.\ 2020, \mnras, 493, 1044. 

% U
% V
\bibitem[van der Walt et al., (2019)]{vdwalt2019} van der Walt et al., (2019) JOSS, 4(37), 1247, 

\bibitem[Vincenzi et al.(2024)]{vincenzi2024} Vincenzi, M., Brout, D., Armstrong, P., et al.\ 2024, arXiv:2401.02945. 

% W
\bibitem[Whelan \& Iben(1973)]{whelan73} Whelan, J. \& Iben, I.\ 1973, \apj, 186, 1007. 

\bibitem[Wiseman et al.(2022)]{wiseman2022} Wiseman, P., Vincenzi, M., Sullivan, M., et al.\ 2022, \mnras, 515, 4587. 

% X
% Y
\bibitem[Yao et al.(2019)]{yao2019} Yao, Y., Miller, A.~A., Kulkarni, S.~R., et al.\ 2019, \apj, 886, 152. 

\bibitem[Yaron \& Gal-Yam(2012)]{yarongalyam2012} Yaron, O. \& Gal-Yam, A.\ 2012, \pasp, 124, 668. 

% Z

\end{thebibliography}
\end{document}